\documentclass[final]{IEEEtran}%
\usepackage{amssymb}
\usepackage{amsmath}
\usepackage{amsfonts}
\usepackage[outercaption]{sidecap}
\usepackage{graphicx}%
\usepackage{color}
\usepackage{cite}
\usepackage{balance}%

\begin{document}
%
%\markboth{Submitted to \textit{IEEE Transactions on Communications}}{Submitted to \textit{IEEE Transactions on Wireless Communications}}

%\pubid{0000--0000/00\$00.00~\copyright~2014 IEEE}
\title{Mobile Communication Systems in the Presence of Fading/Shadowing, Noise and Interference}

\author{Petros~S.~Bithas and Athanasios~A.~Rontogiannis,
%\thanks{Manuscript received October X, 2009; revised XXXXXX 00, 2009;
%        accepted XXXXXX 00, 2009. The editor coordinating the review
%        of this paper and approving it for publication is XXXXXX.}%
%\thanks{This research has been co-financed by the European Union (European Social Fund – ESF) and Greek national funds through the Operational Program "Education and Lifelong Learning" of the National Strategic Reference Framework (NSRF) - Research Funding Program: Thales. Investing in knowledge society through the European Social Fund. }%
\thanks{The authors are with the Institute for Astronomy, Astrophysics, Space Applications and Remote Sensing (IAASARS), Metaxa \& Vas. Pavlou Street, Palea Penteli, 15236 Athens, Greece (e-mail: pbithas@space.noa.gr, tronto@noa.gr).}
%\thanks{Publisher Item Identifier XXXXXX.}
}

\maketitle

\begin{abstract}
In this paper, the effects of interference on composite fading environments, where multipath fading coexists with shadowing, are investigated. Based on some mathematical convenient expressions for the sum of squared $\mathcal{K}$-distributed random variables, which are derived for the first time, important statistical metrics for the signal to interference and noise ratio (SINR) are studied for various cases including non identical, identical and fully correlated statistics. Furthermore, our analysis is extended to multi-channel receivers and in particular to selection diversity (SD) receivers, investigating two distinct cases, namely, signal-to-noise ratio based and SINR-based SD. For all scenarios, simplified expressions are also provided for the interference limited case, where the influence of thermal noise is ignored. The derived expressions are used to analyze the performance, in terms of the average bit error probability (ABEP) and the outage probability (OP), of such systems operating over composite fading environments. A high SNR analysis is also presented for the OP and the ABEP, giving a clear physical insight of the system's performance in terms of the diversity and coding gains. The analysis is accompanied by numerical evaluated results, clearly demonstrating the usefulness of the proposed theoretical framework.
\end{abstract}%


\section{Introduction}
\PARstart{I}{n} many recent wireless communication systems, both licensed, e.g., long term evolution (LTE), or unlicensed, e.g., WiFi or bluetooth, a user quite often shares the same channels with other users and thus in the receiver side the signals need to be intelligently separated. This is imperative, since the aggressive frequency reuse that is frequently employed in cellular systems for increasing spectrum efficiency, causes co-channel as well as adjacent channel interference. The co-channel and adjacent channel interference depend on various physical factors, including interferers' spatial distribution, interfering channels fading, the power of the interferers and the wireless communication system considered. Depending upon the fading characteristics as well as the existence or not of multi-channel transmitters/receivers and/or relays, numerous contributions analyzing the effects of \textit{fading} in conjunction with \textit{interference} have been made, e.g., \cite{923061, radaydeh2009snr, ikki2012multihop, ge2011capacity,benevides2011interference} and the references therein.

In terrestrial (indoor or outdoor) and satellite land-mobile systems, the link quality is also affected by slow variations of the mean signal level due to the \textit{shadowing} from terrain, buildings and trees \cite{B:Sim_Alou_Book}. Under these circumstances, where multipath fading coexists with shadowing, the so-called \textit{composite fading/shadowing environment} originates. In the past, this environment has been statistically modeled by using lognormal-based distributions such as Rayleigh-, Nakagami- and Rice-lognormal \cite{B:Sim_Alou_Book,abu1994micro,yu2010outage} and therefore rather mathematically complicated expressions have been derived for the performance analysis on such communication scenarios. In order to facilitate the communication systems performance evaluation in these environments, new families of distributions that accurately model composite fading conditions have been proposed, most notably as the $\mathcal{K}$ and the generalized-$\mathcal{K}$ ($\mathcal{K}_G$) distributions, e.g., \cite{957106,bithas2006performance,al2010approximation}. Based on the mathematical tractability of these new composite fading models, research efforts have been made recently for investigating the influence that interference has to the system's performance, e.g., \cite{kostic2005analytical,trigui2009performance,anastasov2012analytical}. A common practice in all these works is that the research was restricted to interference-limited wireless communication systems and thus only the statistics of the signal-to-interference-ratio (SIR) was studied.\pubidadjcol

In this paper extending this approach, and in order to provide a more complete stochastic analysis framework of the composite fading environment, the effect of thermal noise is also taken into consideration. This is important since thermal noise may be the main source of system performance degradation especially in cases of weak interfering signals. Therefore, assuming such a complete model, our contribution in this paper can be summarized as follows:
\begin{itemize}
    \item mathematical convenient expressions for the sum of squared $\mathcal{K}$-distributed random variables (RV)s are derived for the first time, accompanied by a convergence analysis for the resulting infinite series formulas
    \item based on these expressions, the signal to interference and noise ratio (SINR) statistics are studied for various scenarios including non identical, identical and fully correlated fading/shadowing effects
    \item the analysis is extended to single input multiple output (SIMO) receivers and in particular to selection diversity (SD) receivers that operate in such environments
    \item the derived expressions are used to investigate the system performance in terms of the average bit error probability (ABEP) and the outage probability (OP)
    \item a high SNR analysis is also provided for the ABEP and the OP in the SIMO case, which is then employed to identify the coding and diversity gains of the system under study.
\end{itemize}
For all scenarios studied, simplified expressions are also provided for the interference limited cases, where the influence of thermal noise is ignored, whilst for SIMO diversity systems an investigation of the system's power efficiency is also presented.

The remainder of this paper is organized as follows. The system model is described in Section II. In Section III the statistics of the sum of squared $\mathcal{K}$-distributed RVs are investigated for the three scenarios under consideration, namely non identical statistics, identical statistics and fully correlated statistics. The statistics of the SINR for single input single output (SISO) and SIMO channels are derived in Sections IV and V, respectively. In Section VI the performance analysis of such systems is presented, while performance evaluation results are provided in Section VII. Finally, concluding remarks are given in Section VIII.
\begin{figure}[!t]
\centering
\hspace{2cm}\includegraphics[keepaspectratio,width=3.2in]{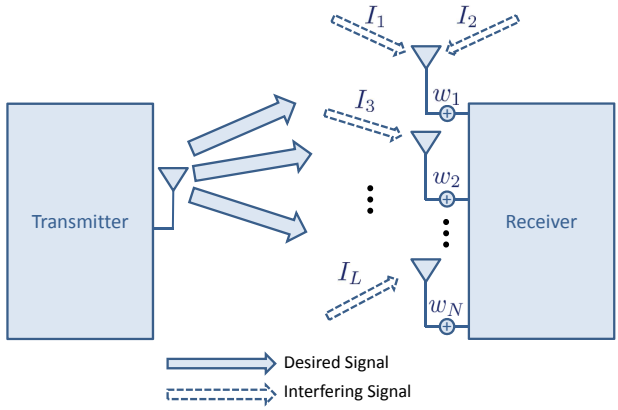}
\caption{System model.} \label{Fig:system_model}
\end{figure}
\section{System Model}\label{sec:system_model}
We consider the downlink of a wireless-mobile communication system, with a single-antenna transmitter and (in general) a multiple-antennas receiver, where each receiver diversity branch is experiencing interference coming from $L$ sources, as shown in Fig.~\ref{Fig:system_model}. We assume that (in general) the level of interference at the receiver is such that the effect of thermal noise on system performance cannot be ignored \cite{923061}. Furthermore, we investigate the case where the desired received signal gain is subject only to multipath fading, while interfering signals are subject to multipath fading and shadowing. Since each user communicates with the base station (BS) providing the highest received SNR, this BS will be with high probability the closest to the user BS. Therefore, not many obstacles between the user and the tagged BS are expected to be present and thus it is reasonable that the received (desired) signal is not subject to shadowing phenomena. On the other hand, due to the expected larger propagation distances, interfering signals are very likely to propagate over obstructed paths, and thus experiencing severe shadowing conditions. Similar assumptions have been made by many other researchers in the past, e.g.,\cite{yao1994cochannel,nakamura1997site,babich1999multimode,radaydeh2009performance,andrews2011tractable}. The complex baseband signal $y_n$ received at the $n$th antenna, can be expressed as
\begin{equation}\label{eq:received_signal}
y_n=h_{D_n}s_D+\sum\limits_{i=1}^L h_{I_{n,i}}s_{I_i} +w_n
\end{equation}
where $h_{D_n}$ represents the complex channel gain between the transmitter and the $n$th receiver antenna (with its envelope following the Rayleigh distribution) and $s_D$ is the desired transmitted complex symbol with energy $E_{s_D}=\mathbb{E}\left< |s_D|^2\right>$, and $\mathbb{E}\left< \cdot\right>$ denoting statistical averaging. Furthermore, in \eqref{eq:received_signal}, $h_{I_{n,i}}$ represents the complex channel gain (with its envelope following the $\mathcal{K}$-distribution) of the interfering signal $s_{I_i}$ and $w_n$ is the complex additive white Gaussian noise (AWGN) with zero mean and variance $N_0$. To proceed, we denote the instantaneous signal-to-noise-ratio (SNR) of the desired signal as $\gamma_{D_n}=|h_{D_n}|^2 E_{s_D}/N_0$, the corresponding average SNR as $\overline{\gamma}_{D_n}=\mathbb{E}\left<|h_{D_n}|^2\right> E_{s_D}/N_0$, whilst the instantaneous interference-to-noise-ratio (INR) of the $i$th interfering signal is defined as $\gamma_{I_{n,i}}=|h_{I_{n,i}}|^2 E_{s_{I_i}}/N_0$, with corresponding average INR equal to $\overline{\gamma}_{I_{n,i}}=\mathbb{E} \left<|h_{I_{n,i}}|^2\right> E_{s_{I_i}}/N_0$.

Since the desired signal is subject only to multipath fading, its instantaneous SNR $\gamma_{D_n}$ at the $n$th antenna can be assumed to be exponentially distributed. Specifically, considering independent fading conditions, the probability density function (PDF) of $\gamma_{D_n}$ is given by
\begin{equation}\label{eq:Rayleigh PDF}
f_{\gamma_{D_n}}(x)= \frac{1}{\overline{\gamma}_{D_n}} \exp \left( -\frac{x}{\overline{\gamma}_{D_n}}\right).
\end{equation}
Here, the instantaneous INR, $\gamma_{I_{n,i}}$, of the interfering signals, which are subject to fading/shadowing effects, is assumed to follow a squared $\mathcal{K}$ distribution\footnote{For simplification purposes and in order to avoid repetitions, from now on squared $\mathcal{K}$ distribution will be referred as $\mathcal{K}$.} with PDF given by \cite{957106}
\begin{equation}\label{eq:d1}
f_{\gamma_{I_{n,i}}}(x)=2\left( \frac{ k_{n,i}}{\overline{\gamma}_{I_{n,i}}}\right)^{\frac{k_{n,i}+1}2}\frac{x^{\frac{k_{n,i}-1}2}}{\Gamma\left( k_{n,i}\right)}
K_{k_{n,i}-1}\left( 2\sqrt{\frac{k_{n,i}x}{\overline{\gamma}_{I_{n,i}}}}\right).
\end{equation}
In \eqref{eq:d1}, $k_{n,i}$ denotes the shaping parameter of the distribution, $\Gamma\left(\cdot\right)$ is the Gamma function \cite[eq. (8.310/1)]{B:Ryzhik_book} and $K_v(\cdot)$ is the second kind modified Bessel function of $v$th order \cite[eq. (8.407/1)]{B:Ryzhik_book}. In addition, $k_{n,i}$ is related to the severity of the shadowing, e.g., for small values for $k_{n,i}$, \eqref{eq:d1} models severe shadowing conditions, while as $k_{n,i}\rightarrow \infty$, it approximates the exponential distribution. The corresponding output SINR, at the $n$th receiver antenna, is expressed as \cite{923061}
\begin{equation}\label{eq:a}
\gamma_{{\rm out}_n}= \frac{\gamma_{D_n}}{ 1+ \gamma_{I_n}}
\end{equation}
where $\gamma_{I_n}$ denotes the total INR\footnote{Without loosing the generality, it is assumed that the all diversity branches are affected by the same interfering sources.}, i.e., $\gamma_{I_n}=\sum\limits_{i=1}^L\gamma_{I_{n,i}}$, and the PDF of $\gamma_{I_{n,i}}$ is given by \eqref{eq:d1}. The PDF of $\gamma_{{\rm out}_n}$ can be evaluated as follows
\begin{equation}\label{eq:b}
f_{\gamma_{{\rm out}_n}}(\gamma) =\int_{0}^\infty \left( 1+x\right)f_{\gamma_{D_n}}((1+x)\gamma)
f_{\gamma_{I_n}}(x)dx.
\end{equation}
In the next section important statistical metrics of $\gamma_{I_n}$ will be evaluated considering non identical, identical as well as fully correlated statistical parameters. The derived results (presented in Section~\ref{sec:sum_of_kappa}), will be used to obtain expressions for the instantaneous output SINR of both SISO and SIMO systems (presented in Sections~\ref{sec:SINR},\ref{SEC:SINR_MB}, respectively).

\section{On the Sum of $\mathcal{K}$-Distributed RVs}\label{sec:sum_of_kappa}
In this section the statistics of the sum of $\mathcal{K}$-distributed RVs will be investigated for three different scenarios, namely for $\mathcal{K}$-distributed RVs which, i) are independent but non identically distributed (i.n.d.), ii) are independent and identically distributed (i.i.d.) and iii) have fully correlated mean values. It should be noted here that in the past, several efforts have been devoted to stochastically characterize such a sum, e.g., \cite{al2010approximation,5501945,atapattu2011mixture}. To the best of the authors knowledge none of them led to exact expressions, which are obtained for the first time in the following analysis. Since the analysis provided in this and the next sections refers to the arbitrary $n$th antenna, the antenna index $n$ will be omitted in this and the next sections, and will be reestablished in Section V.

\newtheorem{theorem}{Theorem}
\newtheorem{lemma}{Lemma}
\subsection{Independent but not Identically Distributed RVs}
\begin{theorem}\label{theor:sum_independent_K_ind}
Let $\gamma_I$ denote a RV defined as
\begin{equation}\label{eq:sum_of_gamma_IND}
\gamma_I\triangleq \sum\limits_{i=1}^{L}\gamma_{I_i}
\end{equation}
where the PDF of $\gamma_{I_i}$ is given by \eqref{eq:d1}. Considering \emph{non identical (interference) statistics}, the PDF of
$\gamma_I$ can be expressed as
\begin{equation}\label{eq:l1}
\begin{split}
 &f_{\gamma_I} (\gamma)=\mathcal{G}_1\left[ \left(\frac{\gamma}{\mathcal{S}_{k_i,\overline{\gamma}_{I_i}}^L}\right) ^{\frac{\mathcal{S}_{k_i,1}^L-1}{2}} I_{\mathcal{S}_{k_i,1}^L-1}\left( 2 \sqrt{\mathcal{S}_{k_i,\overline{\gamma}_{I_i}}^L\gamma} \right) \right.\\ &\left. \hspace{0.3cm}+ \underbrace{\sum_{\substack{i,i\\ \lambda_1,\ldots,\lambda_i}}^L
\sum\limits_{h=0}^\infty \left( \frac{\gamma}{ \mathcal{S}_{k_i,\overline{\gamma}_{I_i}}^L}\right)^{ \frac{\mathcal{G}_{2_i}-1}{2}}\mathcal{G}_3 I_{\mathcal{G}_{2_i}-1} \left(  2 \sqrt{\mathcal{S}_{k_i,\overline{\gamma}_{I_i}}^L\gamma}\right)}_{IS}\right].
\end{split}
\end{equation}
In \eqref{eq:l1}, $\mathcal{G}_{2_j}=\sum\limits_{i=1}^L k_i+j-\sum\limits_{i=1}^jk_{\lambda_i}+h$,
$\mathcal{G}_3=t_{n_{\lambda_1},\lambda _1}$, if the maximum value of the product in \eqref{eq:i1} is equal to 1, $\mathcal{G}_3=\hat{t}_{1,h}=\sum\limits_{p=0}^h t_{p,\lambda _1} t_{h-p,\lambda _2}$, if the maximum value of the product in \eqref{eq:i1} is equal to 2, $\mathcal{G}_3=\hat{t}_{m,h}=\sum\limits_{p=0}^h \hat{t}_{m-1,h} t_{h-p,\lambda _m}$, if the maximum value of the product in \eqref{eq:i1} is $m>2$, with $\hat{t}_{h,i}, t_{h,i}$, $\mathcal{S}_{x_q,y_q}^z$ and $\mathcal{G}_1$ given in Appendix A and $ I_v(\cdot)$ denoting the first kind modified Bessel function of $v$th order \cite[eq. (8.406/1)]{B:Ryzhik_book}.
\end{theorem}

\begin{IEEEproof}
See Appendix~\ref{App:proof_of_theoremI}.
\end{IEEEproof}
It can be shown that the infinite series in $IS$ converges everywhere. Due to space limitations a convergence proof for the general case of $L$ $\mathcal{K}-$distributed RVs is not presented. Instead, the special case of $L=3$ RVs is considered in Appendix~B, where an analytical framework for the convergence of the infinite series in $IS$ can be found. Note that the generalization of this mathematical analysis for $L$ $\mathcal{K}-$distributed RVs is straightforward. In addition the rate of convergence of $f_{\gamma_I}(\gamma)$ given in \eqref{eq:l1}, was investigated experimentally. This is important because in practice, to evaluate $f_{\gamma_I}(\gamma)$, the infinite series in $IS$ is truncated and a number of series terms $H$ is retained. In Table~\ref{Tab:convergence} the minimum values
of $H$, required for accuracy better than $\pm10^{-5}$ are presented for various values of the average INR $\overline{\gamma}_{I_i}$ and distribution's parameters. These results have been obtained by assuming an exponentially decaying profile, i.e., $\overline{\gamma}_{I_i}=10^{{\rm dB}/10}\exp\left[-d (i-1)\right], k_i=3-0.3i$ (with $i=1,2,\ldots,L$), $d=0.1$ and $L=3$. It is clearly depicted from this table that a relatively small number of terms is sufficient to achieve a high accuracy and as a consequence the PDF in \eqref{eq:l1} converges fast. It is also shown that the number of these terms increases as $\overline{\gamma}_{I_i}$ decreases and/or $k_i, \gamma$ increase.

\begin{table}
\renewcommand{\arraystretch}{1}
\caption{Minimum Number of Terms $H$ of
\eqref{eq:l1} Required for Obtaining Accuracy Better Than $10^{\pm5}$.} \label{Tab:convergence}
\centering
\begin{tabular}{||c||c||c||}
  \hline\hline
   & $\bf 5${\bf dB} & $\bf 15${\bf dB}   \\
    \begin{tabular}{c}
    \hline \hline
    ${\bf{ \gamma(\gamma_I)}}$  \\
    \hline
    {\bf 1} \\
    {\bf 5} \\
    {\bf 10} \\
    {\bf 15} \\
    {\bf 20} \\
    \end{tabular}
             & \begin{tabular}{c|c}
            \hline \hline
            $\bf k=1.5$  &  $\bf k=3$  \\
             \hline
                          4 & 5 \\
                         9 & 16 \\
                         13 & 20 \\
                        16 & 23 \\
                        18 & 25 \\
                 \end{tabular}

                                                & \begin{tabular}{c|c}
                                                \hline \hline
                                                $\bf k=1.5$ &  $\bf k=3$   \\
                                                \hline
                         1 & 2 \\
                         4 & 7 \\
                         4 & 8 \\
                        5 & 9 \\
                        7 & 11 \\
                        \end{tabular}\\
\hline \hline
\end{tabular}
\end{table}
\begin{lemma}\label{lemma:CDF_sum_independent_K_iid}
For i.n.d. statistics, the CDF of $\gamma_I$ can be obtained as
\begin{equation}\label{eq:CDF_ind_sum_k}
\begin{split}
  &F_{\gamma_I} (\gamma)= \mathcal{G}_1\left[ \left(\frac{\gamma}{ \mathcal{S}_{k_i,\overline{\gamma}_{I_i}}^L}\right)^{\mathcal{S}_{k_i,2}^L} I_{\mathcal{S}_{k_i,1}^L}\left( 2 \sqrt{\mathcal{S}_{k_i,\overline{\gamma}_{I_i}}^L} \gamma^{1/2}\right) \right. \\ & \left.\hspace{0.3cm}+ \sum_{\substack{i,i\\ \lambda_1,\ldots,\lambda_i}}^L
\sum\limits_{h=0}^\infty \left(\frac{\gamma}{\mathcal{S}_{k_i,\overline{\gamma}_{I_i}}^L}\right) ^{\frac{\mathcal{G}_{2_i}}{2}}\mathcal{G}_3 I_{\mathcal{G}_{2_i}} \left(  2 \sqrt{\mathcal{S}_{k_i,\overline{\gamma}_{I_i}}^L} \gamma^{1/2}\right)\right].
\end{split}
\end{equation}
\end{lemma}

\begin{IEEEproof}
Substituting \eqref{eq:l1} in the definition of the CDF, $F_{\gamma_I} (\gamma)=\int_0^\gamma f_{\gamma_I} (x)dx$, making changes of variables of the form $x^{1/2}=y$ and $y=\sqrt{\gamma}z$ and using
\cite[eq. (6.561/7)]{B:Ryzhik_book}, \eqref{eq:CDF_ind_sum_k} is
obtained.
\end{IEEEproof}

\begin{figure}
\centering\vspace{2.3cm}
\includegraphics[keepaspectratio,width=3in]{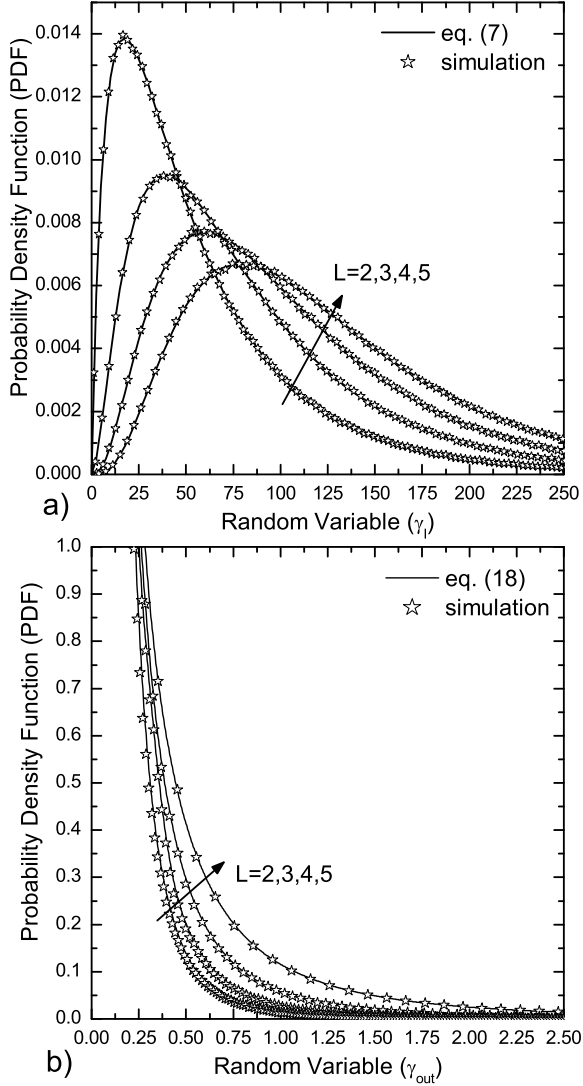}
\caption{ a) The PDF of $\gamma_{I}$ given in \eqref{eq:l1}. b) The PDF of $\gamma_{\rm out}$ given in \eqref{eq:SINR_ind_PDF_final}.} \label{Fig:Fig1}
\end{figure}

In Fig.~\ref{Fig:Fig1}a, several PDFs, obtained by using \eqref{eq:l1}, are plotted for various values of the number of interferers $L$. These distributions have been evaluated by also assuming exponential decaying average INR, i.e., $\overline{\gamma}_{I_i}=10^{{\rm dB}/10}\exp\left[-d (i-1)\right], k_i=k-0.3i$ (with $i=1,2,\ldots,L$), and dB$=15, d=0.1$. Additionally, simulated results are also included in this figure, verifying in all cases the agreement between the analytical and the simulated PDFs.

\subsection{Independent and Identically Distributed RVs}
\begin{theorem}\label{theor:sum_independent_K_iid}
Let $\gamma_I$ denote a RV defined as
\begin{equation}
\gamma_I\triangleq \sum\limits_{i=1}^{L}\gamma_{I_i}
\end{equation}
where the PDF of $\gamma_{I_i}$ is given by \eqref{eq:d1}, with $\overline{\gamma}_{I_i}=\overline{\gamma}_{I}$ and $k_i=k$. Considering \emph{identical (interference) statistics}, the PDF of $\gamma_I$ can be expressed as
\begin{equation}\label{eq:i2}
\begin{split}
&f_{\gamma_I}\left(\gamma\right) =  L\,\sum\limits_{i=0}^L \binom{L}{i} \Gamma \left(1-k \right)^{L-i} (-1)^{i} \\ &  \times \left[ \sum\limits_{h=0}^\infty c_h \left( \frac{k}{\overline{\gamma}_IL}\right)^{\frac{1+\mathcal{G}_4}2} \gamma^{\frac{\mathcal{G}_4-1}2} I_{\mathcal{G}_4-1} \left( 2 \sqrt{\frac{kL}{\overline{\gamma}_I}} \gamma^{1/2}\right) \right]
\end{split}
\end{equation}
where $c_0=a_0^i,\,c_h = \frac1{h a_0} \left[ \sum\limits_{t=1}^h (t i-h+t)a_t c_{h-t} \right] \, {\rm for}\, h\geq1,\, a_0=\frac1{1-k},\,a_t=\frac{(-1)^t}{t! (1-k+t)}$, with $\mathcal{G}_4=L k +h+(1-k)i$.
\end{theorem}

\begin{IEEEproof}
See Appendix~\ref{App:proof_of_theoremII}.
\end{IEEEproof}

\begin{lemma}\label{lemma:CDF_sum_independent_K_iid}
For i.i.d. statistics, the CDF of $\gamma_I$ can be obtained as
\begin{equation}\label{eq:CDF_sum_K_iid}
\begin{split}
&F_{\gamma_I}\left(\gamma\right) =  \sum\limits_{i=0}^L \binom{L}{i} \Gamma \left(1-k \right)^{L-i} (-1)^{i} \\ & \times \left[ \sum\limits_{h=0}^\infty c_h \left( \frac{k}{\overline{\gamma}_I L}\right)^{\frac{\mathcal{G}_4}2}  \gamma^{\frac{\mathcal{G}_4}2} I_{\mathcal{G}_4} \left( 2 \sqrt{\frac{kL}{\overline{\gamma}_I}} \gamma^{1/2}\right) \right].
\end{split}
\end{equation}
\end{lemma}

\begin{IEEEproof}
Substituting \eqref{eq:i2} in the definition of the CDF, $F_{\gamma_I} (\gamma)=\int_0^\gamma f_{\gamma_I} (x)dx$, making changes of variables of the form $x^{1/2}=y$ and $y=\sqrt{\gamma}z$ and using
\cite[eq. (6.561/7)]{B:Ryzhik_book}, \eqref{eq:CDF_ind_sum_k} is
obtained.
\end{IEEEproof}

\subsection{Fully Correlated Shadowing}
When the $L$ different interfering paths exhibit identical shadowing effects, the mean values of the corresponding $\mathcal{K}$-distributed RVs are fully correlated. This is the so-called \textit{fully (or totally) correlated shadowing} communication scenario, which has gained considerable interest lately, e.g., \cite{rosa2010performance,5441355,5599549,al2012asymptotic}, and will be investigated in this section. This type of shadowing arises in situations where the interferers have approximately the same distance from the receiver, and thus the same obstacles shadow in a quite similar way the various interfering signals. As a result, the local mean powers of the interfering signals become correlated \cite{trigui2009performance, wang1996incoherent}, a situation that quite often arises in indoor communication systems \cite{abdel1999performance}.

In such a communication scenario, where the multipath Rayleigh components of the interferers are independent but all of them experience a common local average power, $\overline{\gamma}_I$, it has been shown that the moments generating function (MGF) of $\gamma_I$ can be expressed as \cite[eq. (9)]{957106}
\begin{equation}\label{eq:fully_correlated_MGF}
\mathcal{M}_{\gamma_I}(s)=\left( \frac{k}{\overline{\gamma}_I s}\right)^{k} U\left(k,k-L+1,\frac{k}{\overline{\gamma}_I s} \right)
\end{equation}
where $U(\cdot)$ denotes the confluent hypergeometric function defined as $
U\left(k,k-L+1,\frac{k}{\overline{\gamma}_I s} \right) =\frac{\Gamma\left( L-k\right)}{\Gamma(L)} \null_1F_1 \left( k;k-L+1;\frac{k}{\overline{\gamma}_Is}\right)+\frac{\Gamma\left( k-L\right)}{\Gamma(k)} \left( \frac{k}{\overline{\gamma}_Is}\right)^{L-k}\null_1F_1 \left( L;1-k+L;\frac{k}{\overline{\gamma}_Is}\right)$ \cite[eq. (9.210/2)]{B:Ryzhik_book}, with $\null_1F_1(\cdot)$ representing the confluent hypergeometric function given by \cite[eq. (9.210/1)]{B:Ryzhik_book}. Based on this expression and employing the infinite series representation for $\null_1F_1(\cdot)$, i.e., \cite[eq. (07.20.02.0001.01)]{E:wolfram}, \eqref{eq:fully_correlated_MGF} yields
\begin{equation}\label{eq:fully_correlated_MGF_infinite}
\begin{split}
\mathcal{M}_{\gamma_I}(s)&= \frac{\Gamma(L-k)}{\Gamma(L)} \left( \frac{k}{\overline{\gamma}_Is}\right)^k  \sum\limits_{i=0}^\infty \frac{(k)_i \left[ k/\left(\overline{\gamma}_Is\right)\right]^i}{i!\left(k-L+1 \right)_i } \\ &+\frac{\Gamma(k-L)}{\Gamma(k)} \left( \frac{k}{\overline{\gamma}_Is}\right)^L \sum\limits_{i=0}^\infty \frac{(L)_i \left[ k/\left(\overline{\gamma}_Is\right)\right]^i}{i!\left(1-k+L \right)_i }
\end{split}
\end{equation}
where $(\cdot)_v$ denotes the pochhammer symbol \cite[pp. xliii]{B:Ryzhik_book}. Applying the inverse Laplace transform in \eqref{eq:fully_correlated_MGF_infinite}, yields the following expression for the PDF of $\gamma_I$
\begin{equation}\label{eq:fully_correlated_PDF_infinite}
\begin{split}
f_{\gamma_I}(\gamma)&= \frac{\Gamma(L-k)}{\Gamma(k)\Gamma(L)} \left( \frac{k \,\gamma}{\overline{\gamma}_I}\right)^k \frac1{\gamma} \sum\limits_{i=0}^\infty \frac{(k \,\gamma/\overline{\gamma}_I)^i }{i!\left(k-L+1 \right)_i } \\ &+\frac{\Gamma(k-L)}{\Gamma(k)\Gamma(L)} \left( \frac{k \,\gamma}{\overline{\gamma}_I}\right)^L \frac1{\gamma}\sum\limits_{i=0}^\infty \frac{(k \,\gamma/\overline{\gamma}_I)^i }{i!\left(1-k+L \right)_i }.
\end{split}
\end{equation}
Furthermore, based on the definition of the generalized hypergeometric function, i.e., $ \null_0F_1 \left( b; z\right)= \sum\limits_{i=0}^\infty z^i /\left(i!\left(b\right)_i \right)$ \cite[eq. (07.17.02.0001.01)]{E:wolfram}, and by employing \cite[eq. (03.04.27.0005.01)]{E:wolfram}, a simpler closed-form expression for the PDF of $\gamma_I$, when fully correlated shadowing effects are present, can be extracted as
\begin{equation}\label{eq:fully_correlated_PDF_final}
f_{\gamma_I}(\gamma)=2 \frac{\left( k/\overline{\gamma}_I\right)^{\frac{L+k}2} \gamma^{\frac{L+k}2-1}}{\Gamma(L)\Gamma(k)} K_{L-k}\left( 2 \sqrt{\frac{k}{\overline{\gamma}_I}\gamma}\right).
\end{equation}
Note that in a different research topic, i.e., free space optical communications, a similar PDF expression, as the one given in \eqref{eq:fully_correlated_PDF_final}, has been independently derived in \cite{samimi2012distribution}. Substituting now \eqref{eq:fully_correlated_PDF_final} in the definition of the CDF \cite[eq. (4.17)]{B:Papoulis_book}, using the Meijer-G function representation for the $K_{L-k}(\cdot)$, i.e., \cite[eq. (03.04.26.0006.01)]{E:wolfram}, and then \cite[eq. (26)]{C:Adamchik_Meijer}, the CDF of $\gamma_I$ can be expressed in a simple closed form as
\begin{equation}\label{eq:fully_correlated_CDF_final}
F_{\gamma_I}\left(\gamma\right)=\frac{\left( k/\overline{\gamma}_I\right)^{\frac{L+k}2} }{\Gamma(L)\Gamma(k)}\gamma^{\frac{L+k}2}\mathcal{G} \substack{2,1
\\ 1,3} \left(  \frac{k}{ \overline{\gamma}_I}\gamma
\bigg| \substack{1-\frac{L+k}{2}
\\ \null \\
\frac{L-k}{2},-\frac{L-k}{2},-\frac{L+k}{2}}\right)
\end{equation}
where $\mathcal{G} \substack{m,n \\ p,q} \left[ \cdot | \cdot
\right]$ is the Meijer's $G$-function \cite[eq.
(9.301)]{B:Ryzhik_book}.

It should be noted that the previously derived expressions for the sum of $\mathcal{K}$-distributed RVs can be directly applied to various research topics, different from that treated in this paper, including maximal ratio combiner output SNR study or the analysis of received SNR of cooperative synchronized transmissions \cite{chau2013opportunism}.

\section{SINR Statistics for SISO Systems}\label{sec:SINR}
In this section, based on the previously derived expressions for the sum of $\mathcal{K}$-distributed RVs, a statistical analysis of the instantaneous output SINR $\gamma_{\rm out}$ is presented for the three communication scenarios under consideration, namely i.n.d., i.i.d. and fully correlated interference conditions. In all cases simplified expressions for the SIR are also obtained.

\subsection{Non Identical Interference Statistics}
Substituting the PDF for the SNR given by \eqref{eq:Rayleigh PDF} and the PDF for the INR given by \eqref{eq:l1} in \eqref{eq:b}, it can be shown that integrals of the following form appear
\begin{equation}
\mathcal{I}_1=\int_0^\infty\Phi\left( q_1,\sum\limits_{i=1}^L \frac{k_i}{\overline{\gamma}_{I_i}}\right)dx
\end{equation}
where $\Phi\left(a,b \right)=  x^{\frac{a-1}2} (1+x)\exp \left[ -\frac{\left(1+x\right)\gamma}{\overline{\gamma}_D}\right] I_{a-1} \left(2 \sqrt{b} x^{1/2}\right)$
with $q_1=\sum_{i=1}^Lk_i$ or $q_1=\mathcal{G}_{2_h}$. This type of integrals can be solved in closed form using \cite[eq. (6.643/2)]{B:Ryzhik_book}. After some straightforward mathematical manipulations the PDF of $\gamma_{\rm out}$ under i.n.d. interference conditions can be expressed as
\begin{equation}\label{eq:SINR_ind_PDF_final}
\begin{split}
&f_{\gamma_{\rm out}} (\gamma)= \frac{\exp \left( -\gamma/\overline{\gamma}_D\right)}{\overline{\gamma}_D} \mathcal{G}_1 \exp \left( \frac{\mathcal{S}_{k_i,\overline{\gamma}_{I_i}}^L}{2 \gamma/\overline{\gamma}_D}\right) \left\{ \left( \frac{\overline{\gamma}_D}{\gamma\mathcal{S}_{k_i,\overline{\gamma}_{I_i}}^L}\right)^{\mathcal{S}_{k_i,2}^L} \right. \\ & \left. \times \left[ \mathcal{S}_{k_i,1}^L \frac{\overline{\gamma}_D}{\gamma} M_{-\mathcal{S}_{k_i,2}^L-1,\frac{\mathcal{S}_{k_i,1}^L-1}{2}}\left( \frac{\mathcal{S}_{k_i,\overline{\gamma}_{I_i}}^L} {\gamma/\overline{\gamma}_D}\right)\right. \right.\\ & \left. \left.+
M_{-\mathcal{S}_{k_i,2}^L,\frac{\mathcal{S}_{k_i,1}^L-1}{2}}\left( \frac{\mathcal{S}_{k_i,\overline{\gamma}_{I_i}}^L}{\gamma\overline{\gamma}_D} \right)\right]+ \sum_{\substack{i,i\\ \lambda_1,\ldots,\lambda_i}}^L
\sum\limits_{h=0}^\infty \mathcal{G}_3  \right. \\ & \left. \times \left( \frac{\mathcal{S}_{k_i,\overline{\gamma}_{I_i}}^L\gamma}{\overline{\gamma}_D}\right)^{-\frac{\mathcal{G}_{2_i}}{2}}   \left[  \frac{\mathcal{G}_{2_i}\overline{\gamma}_D}{\gamma}  M_{-\frac{\mathcal{G}_{2_i}}{2}-1,\frac{\mathcal{G}_{2_i}-1}{2}}\left( \frac{\mathcal{S}_{k_i,\overline{\gamma}_{I_i}}^L}{\gamma/\overline{\gamma}_D}\right) \right. \right. \\ & +\left. \left.
M_{-\frac{\mathcal{G}_{2_i}}{2},\frac{\mathcal{G}_{2_i}-1}{2}}\left( \frac{\mathcal{S}_{k_i,\overline{\gamma}_{I_i}}^L}{\gamma/\overline{\gamma}_D}\right)\right]\right\}
\end{split}
\end{equation}
where $M_{\lambda,\mu}(\cdot)$ is the Whittaker function \cite[eq. (9.220/2)]{B:Ryzhik_book}, which is a built-in function in many mathematical software packages. Here it should be mentioned that \eqref{eq:SINR_ind_PDF_final} represents a valid PDF, since it is a nonnegative function, and using \cite[eqs. (07.44.26.0007.01) and (01.03.26.0004.01)]{E:wolfram} and
\cite[eq. (21)]{C:Adamchik_Meijer}, it can be verified that $\int_0^\infty f_{\gamma_{\rm out}} (\gamma)d\gamma=1$ \cite{karagiannidis2006bounds}.
In Fig.~\ref{Fig:Fig1}b, several PDFs, obtained by using \eqref{eq:SINR_ind_PDF_final}, are plotted for the same parameters as the ones used in Fig.~\ref{Fig:Fig1}a and various values for the number of interferers $L$. The simulated results also included in this figure depict in all cases the tight agreement between analytical and simulated PDFs. Furthermore, for the infinite series appearing in \eqref{eq:SINR_ind_PDF_final}, a similar rate of convergence has been observed with that of the series appearing in \eqref{eq:l1}. By definition, the CDF of the instantaneous output SINR is expressed as
\begin{equation}\label{eq:c}
F_{\gamma_{\rm out}}(\gamma) =\int_{0}^\infty F_{\gamma_{D}}((1+x)\gamma)
f_{\gamma_I}(x)dx.
\end{equation}
Substituting the exponential CDF, i.e.,
\begin{equation}\label{eq:Rayleigh_CDF}
F_{\gamma_{D}}(x)=1-\exp \left( -\frac{x}{\overline{\gamma}_D}\right)
\end{equation}
and \eqref{eq:l1} in \eqref{eq:c}, following a similar procedure as the one used for deriving \eqref{eq:SINR_ind_PDF_final}, yields the following closed-form expression for the CDF of $\gamma_{\rm out}$
\begin{equation}\label{eq:SINR_ind_CDF_final}
\begin{split}
&F_{\gamma_{\rm out}} (\gamma)= 1-\exp \left( -\frac{\gamma}{\overline{\gamma}_D}\right) \mathcal{G}_1  \exp \left( \frac{\mathcal{S}_{k_i,\overline{\gamma}_{I_i}}^L}{2 \gamma/\overline{\gamma}_D}\right) \\ & \times  \left\{ \left( \frac{\gamma\mathcal{S}_{k_i,\overline{\gamma}_{I_i}}^L} {\overline{\gamma}_D}\right)^{-\mathcal{S}_{k_i,2}^L}
M_{-\mathcal{S}_{k_i,2}^L,\frac{\mathcal{S}_{k_i,1}^L-1}{2}}\left( \frac{\mathcal{S}_{k_i,\overline{\gamma}_{I_i}}^L}{\gamma\overline{\gamma}_D} \right)\right.\\ &\left. + \sum_{\substack{i,i\\ \lambda_1,\ldots,\lambda_i}}^L
\sum\limits_{h=0}^\infty \mathcal{G}_3  \left( \frac{\mathcal{S}_{k_i,\overline{\gamma}_{I_i}}^L\gamma}{\overline{\gamma}_D}\right)^{-\frac{\mathcal{G}_{2_i}}{2}}   M_{-\frac{\mathcal{G}_{2_i}}{2},\frac{\mathcal{G}_{2_i}-1}{2}}\left( \frac{\mathcal{S}_{k_i,\overline{\gamma}_{I_i}}^L}{\gamma/\overline{\gamma}_D}\right)\right\}.
\end{split}
\end{equation}
For the special case of $L=1$ $\mathcal{K}$-distributed interferer, \eqref{eq:SINR_ind_CDF_final} simplifies to
\begin{equation}\label{eq:CDF_one_interferer}
\begin{split}
 F_{\gamma_{\rm out}}\left(\gamma\right)& = 1-\exp \left( -\frac{\gamma}{\overline{\gamma}_D}\right) \left( \frac{k \overline{\gamma}_D}{\overline{\gamma}_I \gamma}\right)^{\frac{k}{2}}  \\ & \times \exp \left( \frac{k \overline{\gamma}_D}{2 \overline{\gamma}_I \gamma}\right) W_{-\frac{k}{2},\frac{1-k}{2}}\left( \frac{k \overline{\gamma}_D}{\overline{\gamma}_I \gamma}\right).
\end{split}
\end{equation}
In a similar communication scenario consisting of a Nakagami-$m$ desired signal and a Nakagami-lognormal interfering signal, an integral expression for the outage performance was derived in \cite[eq.(3.59)]{B:stuber_Book}. Setting in that expression $m=1$, i.e., considering Rayleigh/Rayleigh-lognormal fading/shadowing conditions, and comparing the resulting formula with \eqref{eq:CDF_one_interferer}, the mathematical simplification offered by the latter is obvious.

\begin{table*}
\renewcommand{\arraystretch}{0.7}
\caption{SISO SIR Statistics.}
\label{Tab:SISO} \centering
\begin{tabular}{c || c }
  \hline\hline
    \begin{tabular}{c}
                \\   Non Identical\\ Interference\\ \\ \\ \\ \\ \\ \\

       \hline   \hline \\ \\
  Identical \\Interference\\ \\ \\ \\
\hline \hline\\
Fully Correlated\\ Interference\\

        \end{tabular}
             & \begin{tabular}{c}
                                 $ f_{\gamma_{\rm out}}(\gamma)= \mathcal{G}_1 \exp \left( \frac{\mathcal{S}_{k_i,\overline{\gamma}_{I_i}}^L}{2 \gamma/\overline{\gamma}_D}\right) \frac{1}{\gamma}\left\{ \left(\frac{\mathcal{S}_{k_i,\overline{\gamma}_{I_i}}^L}{ \frac{\gamma}{\overline{\gamma}_D}}\right)^{\mathcal{S}_{k_i,2}^L} \mathcal{S}_{k_i,1}^LM_{-\mathcal{S}_{k_i,2}^L-1,\frac{\mathcal{S}_{k_i,1}^L-1}{2}}\left( \frac{\mathcal{S}_{k_i,\overline{\gamma}_{I_i}}^L}{\gamma/ \overline{\gamma}_D}\right)\right.$ \\ $\left. + \sum_{\substack{i,i\\ \lambda_1,\ldots,\lambda_i}}^L
                                 \sum\limits_{h=0}^\infty \mathcal{G}_3 \mathcal{G}_{2_i}\left( \frac{\mathcal{S}_{k_i,\overline{\gamma}_{I_i}}^L\gamma}{\overline{\gamma}_D}\right) ^{-\frac{\mathcal{G}_{2_i}}{2}}  M_{-\frac{\mathcal{G}_{2_i}}{2}-1,\frac{\mathcal{G}_{2_i}-1}{2}}\left( \frac{\mathcal{S}_{k_i,\overline{\gamma}_{I_i}}^L}{\gamma/\overline{\gamma}_D}\right)\right\}$ \\
                                 \hline \\
                                 $F_{\gamma_{\rm out}} (\gamma)= 1- \mathcal{G}_1 \exp \left( \frac{\mathcal{S}_{k_i,\overline{\gamma}_{I_i}}^L}{2 \gamma/\overline{\gamma}_D}\right) \left\{ \left( \frac{\mathcal{S}_{k_i,\overline{\gamma}_{I_i}}^L\gamma}{\overline{\gamma}_D}\right)^{-\mathcal{S}_{k_i,2}^L} M_{-\mathcal{S}_{k_i,2}^L,\frac{\mathcal{S}_{k_i,1}^L-1}{2}}\left( \frac{\mathcal{S}_{k_i,\overline{\gamma}_{I_i}}^L}{\gamma\overline{\gamma}_D} \right)\right.$\\$\left.+ \sum_{\substack{i,i\\ \lambda_1,\ldots,\lambda_i}}^L \sum\limits_{h=0}^\infty \mathcal{G}_3    \left( \frac{\mathcal{S}_{k_i,\overline{\gamma}_{I_i}}^L\gamma}{\overline{\gamma}_D}\right)^{-\frac{\mathcal{G}_{2_i}}{2}}   M_{-\frac{\mathcal{G}_{2_i}}{2},\frac{\mathcal{G}_{2_i}-1}{2}}\left( \frac{\mathcal{S}_{k_i,\overline{\gamma}_{I_i}}^L}{\gamma/\overline{\gamma}_D}\right)\right\}$\\

            \hline \hline
                \\   $f_{\gamma_{\rm out}}\left(\gamma\right)  = \sum\limits_{i=0}^L \binom{L}{i} \Gamma \left(1-k \right)^{L-i} (-1)^{i} \sum\limits_{n=0}^\infty c_n \frac{\mathcal{G}_4}{\gamma}   \left( \frac{k \overline{\gamma}_D}{\overline{\gamma}_I L \gamma}\right)^{\frac{\mathcal{G}_4}2}  \exp\left( \frac{kL \overline{\gamma}_D}{2 \overline{\gamma}_I\gamma}\right)
M_{-\frac{\mathcal{G}_4+2}{2}, \frac{\mathcal{G}_4-1}{2}}\left(  \frac{kL \overline{\gamma}_D}{ \overline{\gamma}_I\gamma}\right)$   \\
\hline \\
$F_{\gamma_{\rm out}}\left(\gamma\right) = 1- \sum\limits_{i=0}^L  \binom{L}{i}  \Gamma \left(1-k \right)^{L-i} (-1)^{i}  \sum\limits_{n=0}^\infty c_n \left( \frac{k\overline{\gamma}_D}{\gamma L\overline{\gamma}_I}\right)^{\frac{\mathcal{G}_4}2} \exp \left( \frac{k L\overline{\gamma}_D}{2\gamma\overline{\gamma}_I}\right)
M_{-\frac{\mathcal{G}_4}{2},\frac{\mathcal{G}_4-1}{2}}\left( \frac{k L\overline{\gamma}_D}{\gamma\overline{\gamma}_I}\right)$\\
             \hline \hline
                \\   $f_{\gamma_{\rm out}}\left(\gamma\right)= L k \left( \frac{ k\overline{\gamma}_D}{\overline{\gamma}_I \gamma}\right)^{\frac{L+k-1}{2}} \exp \left( \frac{ k\overline{\gamma}_D}{2\overline{\gamma}_I \gamma}\right)  \gamma^{-1} W_{-\frac{L+k+1}{2},\frac{L-k}{2}}\left( \frac{ k\overline{\gamma}_D}{\overline{\gamma}_I \gamma}\right).$   \\
             \hline \\
             $F_{\gamma_{\rm out}}\left(\gamma\right) =1- \left( \frac{k \overline{\gamma}_D}{\overline{\gamma}_I  \gamma}\right)^{\frac{L+k-1}{2}} \exp \left( \frac{k \overline{\gamma}_D}{2\overline{\gamma}_I  \gamma}\right)   W_{\frac{1-L-k}2,\frac{L-k}{2}} \left( \frac{k \overline{\gamma}_D}{\overline{\gamma}_I  \gamma}\right).$ \\
             \hline \\
             $M_{\gamma_{\rm out}}\left(s\right)=\frac{1}{\Gamma \left( k\right)\Gamma \left( L\right)} \left( \frac{k \overline{\gamma}_D}{\overline{\gamma}_I}\right)^{\frac{L+k-1}{2}} s^{\frac{L+k-1}{2}} \mathcal{G} \substack{1,3
\\ 3,1} \left(  \frac{\overline{\gamma}_I}{ k \overline{\gamma}_D s}
\bigg| \substack{\frac{L+k+1}{2},1-\frac{L-k+1}{2},1+\frac{L-k-1}{2}
\\ \null \\
1+\frac{L+k-1}{2}}\right).$ \\
            \end{tabular}\\
\hline \hline
\end{tabular}
\end{table*}

\textit{\underline{Simplified Expressions for the SIR:}} In many cases, the mobile communication systems tend to be interference limited rather than noise limited, since the thermal and man-made noise effects are often insignificant compared to the signal levels
of cochannel users \cite{andersen1995propagation}. This case will be also studied here, where considering an interference limited environment, i.e., ignoring the AWGN at the user terminal, the received SIR is given by
\begin{equation}\label{eq:SIR_definition}
\gamma_{\rm out}=\frac{ \gamma_D}{ \gamma_I}
\end{equation}
whilst its PDF and CDF expressions are
\begin{equation}\label{eq:SIR_definition}
\begin{split}
f_{\gamma_{\rm out}}(\gamma) &=\int_{0}^\infty x f_{\gamma_D}(x\gamma)
f_{\gamma_I}(x)dx\\
F_{\gamma_{\rm out}}(\gamma) &=\int_{0}^\infty F_{\gamma_D}(x\gamma)
f_{\gamma_I}(x)dx
\end{split}
\end{equation}
respectively. Substituting \eqref{eq:Rayleigh PDF} and \eqref{eq:l1} (or \eqref{eq:Rayleigh_CDF} and \eqref{eq:l1}) in \eqref{eq:SIR_definition}, and following a similar procedure as the one for deriving \eqref{eq:SINR_ind_PDF_final}, yields the PDF and CDF expressions, respectively, for the i.n.d. case given in Table~\ref{Tab:SISO}.
\subsection{Identical Interference Statistics}
Substituting \eqref{eq:Rayleigh PDF} and \eqref{eq:i2} in \eqref{eq:b} and after some mathematical procedure yields
\begin{equation}\label{eq:SINR_iid_integral}
\begin{split}
f_{\gamma_{\rm out}} &\left(\gamma\right) =\frac{L}{\overline{\gamma}_D} \sum\limits_{i=0}^L \binom{L}{i} \Gamma \left(1-k \right)^{L-i} (-1)^{i}  \sum\limits_{n=0}^\infty c_n \\ & \times
\left( \frac{k}{\overline{\gamma}_I L}\right)^{\frac{\mathcal{G}_4}2}\left[\int_0^\infty \Phi \left(\mathcal{G}_4,\frac{kL}{\overline{\gamma}_I} \right) dx\right] .
\end{split}
\end{equation}
The integral in \eqref{eq:SINR_iid_integral} can be solved in closed form using \cite[eq. (6.643/2)]{B:Ryzhik_book} and thus the PDF of $\gamma_{\rm out}$ under i.i.d. interference conditions can be expressed as
\begin{equation}\label{eq:SINR_iid_final}
\begin{split}
& f_{\gamma_{\rm out}}\left(\gamma\right) = \frac{\exp \left( -\gamma/\overline{\gamma}_D\right)}{\overline{\gamma}_D} \sum\limits_{i=0}^L \binom{L}{i}  \Gamma \left(1-k \right)^{L-i} (-1)^{i} \\ & \times \sum\limits_{n=0}^\infty c_n \left( \frac{k\overline{\gamma}_D}{\gamma L\overline{\gamma}_I}\right)^{\frac{\mathcal{G}_4}2} \exp \left( \frac{k L\overline{\gamma}_D}{2\gamma\overline{\gamma}_I}\right)
\left[M_{-\frac{\mathcal{G}_4}{2},\frac{\mathcal{G}_4-1}{2}}\left( \frac{k L\overline{\gamma}_D}{\gamma\overline{\gamma}_I}\right) \right. \\ & \left.+\frac{\mathcal{G}_4\overline{\gamma}_D}{\gamma}M_{-\frac{\mathcal{G}_4+2}{2}, \frac{\mathcal{G}_4-1}{2}}\left( \frac{k L\overline{\gamma}_D}{\gamma\overline{\gamma}_I}\right)\right] .
\end{split}
\end{equation}
The corresponding expression for the CDF is given by
\begin{equation}\label{eq:SINR_iid_final}
\begin{split}
&F_{\gamma_{\rm out}}\left(\gamma\right)= 1-\exp \left( -\frac{\gamma}{\overline{\gamma}_D}\right) \sum\limits_{i=0}^L  \binom{L}{i}  \Gamma \left(1-k \right)^{L-i} (-1)^{i} \\ & \times \sum\limits_{n=0}^\infty c_n \left( \frac{k\overline{\gamma}_D}{\gamma L\overline{\gamma}_I}\right)^{\frac{\mathcal{G}_4}2} \exp \left( \frac{k L\overline{\gamma}_D}{2\gamma\overline{\gamma}_I}\right)
M_{-\frac{\mathcal{G}_4}{2},\frac{\mathcal{G}_4-1}{2}}\left( \frac{k L\overline{\gamma}_D}{\gamma\overline{\gamma}_I}\right) .
\end{split}
\end{equation}

For the SIR case simplified expression for the PDF and CDF are given in Table~\ref{Tab:SISO}.

\subsection{Fully Correlated Interference Statistics}\label{sub:PDF_fully_correlated}
Substituting \eqref{eq:Rayleigh PDF} and \eqref{eq:fully_correlated_PDF_final} in \eqref{eq:b} yields the following expression for the SINR of $f_{\gamma_{\rm out}}\left(\gamma\right)$
\begin{equation}\label{eq:SINR_integral_fully_correlated}
\begin{split}
&f_{\gamma_{\rm out}}\left(\gamma\right)= \frac2{\overline{\gamma}_D} \frac{\left( k/\overline{\gamma}_I\right)^{\frac{L+k}2}}{\Gamma(L)\Gamma(k)} \underbrace{\int_0^\infty x^{\frac{L+k}2-1} (1+x)}_{\mathcal{I}_2} \\ & \times
\underbrace{\exp \left[ -\frac{\left(1+x\right)\gamma}{\overline{\gamma}_D}\right]  K_{L-k} \left( 2 \sqrt{\frac{k}{\overline{\gamma}_I}} x^{1/2}\right)dx}_{\mathcal{I}_2}.
\end{split}
\end{equation}
After performing some straightforward mathematical manipulations and using \cite[eq. (6.643/3)]{B:Ryzhik_book} a closed-form expression for the PDF of $\gamma_{\rm out}$ can be derived as
\begin{equation}\label{eq:SINR_final_fully_correlated}
\begin{split}
&f_{\gamma_{\rm out}}\left(\gamma\right)= \frac1{\overline{\gamma}_D} \left(\frac{k\overline{\gamma}_D}{\gamma \overline{\gamma}_I}\right)^{\frac{L+k-1}{2}} \exp \left( -\frac{\gamma}{\overline{\gamma}_D}\right)\exp \left( \frac{k\overline{\gamma}_D}{\gamma \overline{\gamma}_I}\right)
\\ & \times \left[W_{\frac{1-L-k}{2},\frac{L-k}{2}}\left(\frac{k\overline{\gamma}_D}{\gamma \overline{\gamma}_I} \right) +  \frac{L k\overline{\gamma}_D}{\gamma}W_{-\frac{L+k+1}{2},\frac{L-k}{2}}\left(\frac{k\overline{\gamma}_D}{\gamma \overline{\gamma}_I} \right)\right]
\end{split}
\end{equation}
where $W_{\lambda,\mu}(\cdot)$ is the Whittaker function \cite[eq. (9.220/4)]{B:Ryzhik_book}. Substituting the exponential CDF and \eqref{eq:fully_correlated_PDF_final} in \eqref{eq:c} and using \cite[eqs. (6.561/16 and 6.631/3)]{B:Ryzhik_book} yields the following closed-form expression for the CDF of $\gamma_{\rm out}$
\begin{equation}\label{eq:CDF_SINR_fully_correlated_closed}
\begin{split}
& F_{\gamma_{\rm out}}\left(\gamma\right) = 1-\exp \left( -\frac{\gamma}{\overline{\gamma}_D}\right) \left( \frac{k \overline{\gamma}_D}{\overline{\gamma}_I \gamma}\right)^{\frac{L+k-1}{2}}  \\ & \times \exp \left( \frac{k \overline{\gamma}_D}{2 \overline{\gamma}_I \gamma}\right) W_{\frac{1-L-k}{2},\frac{L-k}{2}}\left( \frac{k \overline{\gamma}_D}{\overline{\gamma}_I \gamma}\right).
\end{split}
\end{equation}

\textit{\underline{Simplified Expressions for the SIR:}} The closed-form expression for the PDF and the CDF of $\gamma_{\rm out}$ are given in Table~\ref{Tab:SISO}.

\section{SINR Statistics for SIMO: The Fully Correlated Case}\label{SEC:SINR_MB}
In this section we consider a SD receiver and investigate two distinct selection techniques, namely the SNR-based and the SINR-based, assuming fully correlated shadowing on the interfering signals. We also assume that the receive antennas are sufficiently spaced, so that the $L$ interfering signals received in any of them are totally independent from the ones received by any other antenna. In this context, new closed-form expressions are derived for important statistical metrics of the instantaneous output SINR of the two techniques under consideration. It is noted that the analytical framework presented here can also be applied to the i.n.d. as well as to i.i.d. interference scenarios. However, due to space limitations these results are not presented here.

\subsection{SNR-Based Criterion}
For the SNR-based SD criterion in the presence of AWGN and multiple interfering signals, the diversity receiver monitors the available diversity branches continuously and selects the branch with the largest instantaneous SNR for data detection. This SD technique requires the separation of the desired signal from the interfering signals, which can be practically achieved by using different pilot signals for each of them \cite{radaydeh2009snr}. Mathematically speaking the instantaneous system output SINR of this SIMO system can be expressed as
\begin{equation}\label{eq:SINR_Diversity}
\gamma_{\rm SD_{out}}= \frac{ \gamma_{\rm SD}}{ 1+  \gamma_{I}}
\end{equation}
where $\gamma_{\rm SD}=\max\{\gamma_{D_1},\gamma_{D_2},\ldots, \gamma_{D_N}\}$ represents the instantaneous output SNR of the SD receiver, with $\gamma_{D_n}$ denoting the instantaneous SNR of the $n$th branch, following the PDF given by \eqref{eq:Rayleigh PDF}. The CDF of $\gamma_{\rm SD}$ for i.n.d. fading conditions\footnote{For SNR-based SIMO, the i.n.d. and i.i.d. conditions refer to the fading statistics of the instantaneous SNRs of the \underline{desired signal} at the branches of the SIMO receiver. For the interfering signals, fully correlated shadowing has been assumed.}, is given by
\begin{equation}\label{eq:CDF_SD_SNR_based_defintion}
F_{\gamma_{\rm SD}} (x)= \prod_{n=1}^N F_{\gamma_{D_n}}(x).
\end{equation}
For i.i.d. fading conditions, \eqref{eq:CDF_SD_SNR_based_defintion} simplifies to $F_{\gamma_{\rm SD}} (x)= \left[ F_{\gamma_D}(x)\right]^N$, with $F_{\gamma_D}(x)$ given in \eqref{eq:Rayleigh_CDF}. Based on \eqref{Eq:product} and after some straightforward mathematical manipulations, $F_{\gamma_{\rm SD}} (x)$ can be expressed as
\begin{equation}\label{eq:CDF_SD_output_IND}
F_{\gamma_{\rm SD}} (x)=1  + \sum_{\substack{n,n\\ \lambda_1,\ldots,\lambda_n}}^N
\exp\left( - \mathcal{S}_{1,\overline{\gamma}_{D_{\lambda_m}}}^n x\right)
\end{equation}
which for the case of i.i.d. fading conditions simplifies to
\begin{equation}\label{eq:CDF_SD_output_IID}
F_{\gamma_{\rm SD}} (x)=\sum\limits_{n=0}^N \binom{N}{n} (-1)^n  \exp \left( -\frac{n}{\overline{\gamma}_D}x\right).
\end{equation}
\begin{table*}
\renewcommand{\arraystretch}{0.8}
\caption{SIMO SIR Statistics.}
\label{Tab:SIMO} \centering
\begin{tabular}{c || c }
  \hline\hline
    \begin{tabular}{c}
               Non Identical \\ Fading\\ \\

       \hline   \hline \\
 Identical\\ Fading
        \end{tabular}
             & \begin{tabular}{c}
                         $ f_{\gamma_{\rm SD_{out}}} (\gamma)=\sum_{\substack{n,n\\ \lambda_1,\ldots,\lambda_n}}^N
                            \frac{L k}{\gamma} \left( \frac{k /\overline{\gamma}_{I}}{ \mathcal{S}_{1,\overline{\gamma}_{D_{\lambda_m}}}^n\gamma}\right)^{\frac{L+k-1}{2}}    \exp \left( \frac{k /\overline{\gamma}_{I}}{ 2\mathcal{S}_{1,\overline{\gamma}_{D_{\lambda_m}}}^n\gamma}\right)  W_{-\frac{L+k+1}{2},\frac{L-k}2} \left( \frac{k /\overline{\gamma}_{I}}{ \mathcal{S}_{1,\overline{\gamma}_{D_{\lambda_m}}}^n\gamma}\right)$ \\
                            \hline \\
                            $F_{\gamma_{\rm SD_{out}}} (\gamma)=1  + \sum_{\substack{n,n\\ \lambda_1,\ldots,\lambda_n}}^N
                            \left( \frac{k /\overline{\gamma}_{I}}{ \mathcal{S}_{1,\overline{\gamma}_{D_{\lambda_m}}}^n\gamma}\right)^{\frac{L+k-1}{2}}  \exp \left( \frac{k /\overline{\gamma}_{I}}{ 2\mathcal{S}_{1,\overline{\gamma}_{D_{\lambda_m}}}^n\gamma}\right) W_{\frac{1-L-k}{2},\frac{L-k}2} \left( \frac{k /\overline{\gamma}_{I}}{ \mathcal{S}_{1,\overline{\gamma}_{D_{\lambda_m}}}^n\gamma}\right)$\\ \hline \\
                            $M_{\gamma_{\rm SD_{out}}} (s)=\sum_{\substack{n,n+1\\ \lambda_1,\ldots,\lambda_n}}^N
                \left( \frac{k /\overline{\gamma}_{I}}{ \mathcal{S}_{1,\overline{\gamma}_{D_{\lambda_m}}}^n}\right)^{ \frac{L+k-1}{2}}\frac{s^{\frac{L+k-3}2}}{\Gamma(k)\Gamma(L)}   \mathcal{G} \substack{1,3 \\ 3,1} \left(  \frac{\frac{\overline{\gamma}_I}{ k}\mathcal{S}_{1,\overline{\gamma}_{D_{\lambda_m}}}^n }{ s}
                \Bigg| \substack{\frac{L+k+1}{2},\frac{1-L+k}{2},\frac{1+L-k}{2}
                \\ \null \\
                \frac{L+k+1}{2}}\right)$\\
            \hline \hline
                \\   $f_{\gamma_{\rm SD_{out}}} (\gamma)=L k\sum\limits_{n=1}^N \binom{N}{n} (-1)^{n+1}  \left( \frac{k \overline{\gamma}_D}{\overline{\gamma}_I n \gamma}\right)^{\frac{L+k-1}{2}}  \exp \left( \frac{k \overline{\gamma}_D}{2\overline{\gamma}_I n \gamma}\right)  \gamma^{-1}W_{\frac{-1-L-k}2,\frac{L-k}{2}} \left( \frac{k \overline{\gamma}_D}{\overline{\gamma}_I n \gamma}\right)$   \\ \hline \\
                $F_{\gamma_{\rm SD_{out}}} (\gamma)=1+\sum\limits_{n=1}^N \binom{N}{n} (-1)^{n}  \left( \frac{k \overline{\gamma}_D}{\overline{\gamma}_I n \gamma}\right)^{\frac{L+k-1}{2}}  \exp \left( \frac{k \overline{\gamma}_D}{2\overline{\gamma}_I n \gamma}\right) W_{\frac{1-L-k}2,\frac{L-k}{2}} \left( \frac{k \overline{\gamma}_D}{\overline{\gamma}_I n \gamma}\right)$\\ \hline \\
                $M_{\gamma_{\rm SD_{out}}} (s)=\sum\limits_{n=1}^N \binom{N}{n} (-1)^{n+1} \left( \frac{k \overline{\gamma}_D}{n\overline{\gamma}_I}\right)^{\frac{L+k-1}{2}}  \frac{s^{\frac{L+k-3}{2}}}{\Gamma(k) \Gamma(L)}  \mathcal{G} \substack{1,3
\\ 3,1} \left(  \frac{n \overline{\gamma}_I }{  k \overline{\gamma}_Ds}
\bigg| \substack{\frac{L+k+1}{2},\frac{k-L+1}{2},\frac{1+L-k}{2}
\\ \null \\
\frac{L+k+1}{2}}\right)$
            \end{tabular}\\
\hline \hline
\end{tabular}
\end{table*}

Starting from \eqref{eq:SINR_Diversity} and substituting \eqref{eq:CDF_SD_output_IND} and \eqref{eq:fully_correlated_PDF_final} in \eqref{eq:c}, integrals of the form $\mathcal{I}_2$ appearing in \eqref{eq:SINR_integral_fully_correlated}, need to be solved. Therefore following the procedure proposed in \ref{sub:PDF_fully_correlated}, the CDF of $\gamma_{\rm SD_{out}}$, with SNR-based SD, can be obtained in closed form as
\begin{equation}\label{eq:CDF_SNR_based_with_SD_IND}
\begin{split}
&F_{\gamma_{\rm SD_{out}}}(\gamma)\hspace{-0.1cm}=\hspace{-0.1cm}1 + \sum_{\substack{n,n\\ \lambda_1,\ldots,\lambda_n}}^N
\hspace{-0.15cm}\left( \frac{k /\overline{\gamma}_{I}}{ \mathcal{S}_{1,\overline{\gamma}_{D_{\lambda_m}}}^n\gamma}\right)^{\frac{L+k-1}{2}} \hspace{-0.1cm}\exp \left(- \mathcal{S}_{1,\overline{\gamma}_{D_{\lambda_m}}}^n\gamma\right) \\ & \times \exp \left( \frac{k /\overline{\gamma}_{I}}{ 2\mathcal{S}_{1,\overline{\gamma}_{D_{\lambda_m}}}^n\gamma}\right) W_{\frac{1-L-k}{2},\frac{L-k}2} \left( \frac{k /\overline{\gamma}_{I}}{ \mathcal{S}_{1,\overline{\gamma}_{D_{\lambda_m}}}^n\gamma}\right).
\end{split}
\end{equation}
Its corresponding PDF is given by
\begin{equation}\label{eq:PDF_SNR_based_with_SD_IND}
\begin{split}
&f_{\gamma_{\rm SD_{out}}} (\gamma)=\sum_{\substack{n,n\\ \lambda_1,\ldots,\lambda_n}}^N
\left( - \mathcal{S}_{1,\overline{\gamma}_{D_{\lambda_m}}}^n\right)  \exp \left(- \mathcal{S}_{1,\overline{\gamma}_{D_{\lambda_m}}}^n\gamma\right)  \\ & \times \left( \frac{k /\overline{\gamma}_{I}}{ \mathcal{S}_{1,\overline{\gamma}_{D_{\lambda_m}}}^n\gamma}\right)^{\frac{L+k-1}{2}} \exp \left( \frac{k /\overline{\gamma}_{I}}{ 2\mathcal{S}_{1,\overline{\gamma}_{D_{\lambda_m}}}^n\gamma}\right) \\ & \times \left[   W_{\frac{1-L-k}{2},\frac{L-k}2} \left( \frac{k /\overline{\gamma}_{I}}{ \mathcal{S}_{1,\overline{\gamma}_{D_{\lambda_m}}}^n\gamma}\right) \right. \\ & \left.+\frac{L k }{\mathcal{S}_{1,\overline{\gamma}_{D_{\lambda_m}}}^n\gamma}  W_{-\frac{L+k+1}{2},\frac{L-k}2} \left( \frac{k /\overline{\gamma}_{I}}{ \mathcal{S}_{1,\overline{\gamma}_{D_{\lambda_m}}}^n\gamma}\right)\right].
\end{split}
\end{equation}
Considering i.i.d. fading conditions \eqref{eq:CDF_SNR_based_with_SD_IND} simplifies to
\begin{equation}\label{eq:eq:CDF_SNR_based_with_SD_IID}
\begin{split}
&F_{\gamma_{\rm SD_{out}}} (\gamma)=1-\sum\limits_{n=1}^N\binom{N}{n} (-1)^{n+1} \exp \left( -\frac{n}{\overline{\gamma}_D}\gamma\right)  \\ & \times \left( \frac{k \overline{\gamma}_D}{\overline{\gamma}_I n \gamma}\right)^{\frac{L+k-1}{2}} \exp \left( \frac{k \overline{\gamma}_D}{2\overline{\gamma}_I n \gamma}\right) W_{\frac{1-L-k}2,\frac{L-k}{2}} \left( \frac{k \overline{\gamma}_D}{\overline{\gamma}_I n \gamma}\right)
\end{split}
\end{equation}
whilst \eqref{eq:PDF_SNR_based_with_SD_IND} simplifies to
\begin{equation}\label{eq:eq:PDF_SNR_based_with_SD_IID}
\begin{split}
&f_{\gamma_{\rm SD_{out}}} (\gamma)=\sum\limits_{n=1}^N \binom{N}{n} (-1)^{n} \left( -\frac{n}{\overline{\gamma}_D}\right)\exp \left( -\frac{n}{\overline{\gamma}_D}\gamma\right) \\ & \times\left( \frac{k \overline{\gamma}_D}{\overline{\gamma}_I n \gamma}\right)^{\frac{L+k-1}{2}} \exp \left( \frac{k \overline{\gamma}_D}{2\overline{\gamma}_I n \gamma}\right)  \left\{ W_{\frac{1-L-k}2,\frac{L-k}{2}} \left( \frac{k \overline{\gamma}_D}{\overline{\gamma}_I n \gamma}\right)\right. \\ & \left. +L k \left( \frac{n}{\overline{\gamma}_D}\gamma\right)^{-1}W_{\frac{-1-L-k}2,\frac{L-k}{2}} \left( \frac{k \overline{\gamma}_D}{\overline{\gamma}_I n \gamma}\right) \right\}.
\end{split}
\end{equation}
\textit{\underline{Simplified Expressions for the SIR:}} For the SIR case, the instantaneous system output SIR can be expressed as
\begin{equation}\label{eq:SIR_Diversity}
\gamma_{\rm SD_{out}}= \frac{ \gamma_{\rm SD}}{  \gamma_I}.
\end{equation}
The simplified PDF and CDF expressions for both i.n.d. and i.i.d. cases are given in Table~\ref{Tab:SIMO}.

\subsection{SINR-Based Criterion}
Under a SINR-based criterion, the diversity receiver selects the branch with the highest instantaneous SINR for coherent detection. This technique is more complex to be implemented, than the SNR-based one, since computationally demanding processing operations are required at the receiving end \cite{radaydeh2009snr}. These include signal separation, SINR calculation per branch, statistical ordering of the resulting SINRs and SINR-based selection via a maximum selection criterion.

For the SINR-based scenario the instantaneous SINR of the system output can be expressed as $\gamma_{\rm SD_{out}}=\max(\gamma_{{\rm out}_1},\gamma_{{\rm out}_2},\ldots,\gamma_{{\rm out}_N})$, where $\gamma_{{\rm out}_n}$ represents the instantaneous SINR of the $n$th branch, with PDF and CDF given by \eqref{eq:SINR_final_fully_correlated} and \eqref{eq:CDF_SINR_fully_correlated_closed}, respectively. Therefore, considering i.n.d. fading conditions\footnote{For the SINR-based criterion the i.n.d. and i.i.d. conditions refer to the fading statistics of the instantaneous SINRs $\gamma_{{\rm out}_n}$.} the CDF of the output SINR can be expressed as
\begin{equation}\label{eq:CDF_ind_SINR-SD}
F_{\gamma_{\rm SD_{out}}}\left(\gamma\right) = \left[\prod_{n=1}^NF_{\gamma_{{\rm out}_n}} (\gamma)\right]
\end{equation}
with $F_{\gamma_{{\rm out}_n}}(\cdot)$ given by \eqref{eq:CDF_SINR_fully_correlated_closed}. For the i.i.d. fading case, \eqref{eq:CDF_ind_SINR-SD} simplifies to the following expression
\begin{equation}\label{eq:CDF_iid_SINR-SD}
F_{\gamma_{\rm SD_{out}}}\left(\gamma\right) = \left[F_{\gamma_{\rm out}}(\gamma)\right]^N.
\end{equation}
The corresponding expressions for the PDFs, considering i.n.d. fading conditions, are
\begin{equation}\label{eq:PDF_SINR_SD_ind}
f_{\gamma_{\rm SD_{out}}}(\gamma)=\sum\limits_{n=1}^Nf_{\gamma_{{\rm out}_n}}(\gamma) \prod_{\substack{m=1\\m\neq n}}^N F_{\gamma_{{\rm out}_m}}(\gamma)
\end{equation}
with $f_{\gamma_{{\rm out}_n}}(\cdot)$ given by \eqref{eq:SINR_final_fully_correlated}. For i.i.d. fading conditions \eqref{eq:PDF_SINR_SD_ind} simplifies to
\begin{equation}\label{eq:PDF_SINR_SD_iid}
f_{\gamma_{\rm SD_{out}}}(\gamma)=N f_{\gamma_{\rm out}}(\gamma) \left( F_{\gamma_{\rm out}}(\gamma)\right)^{N-1}.
\end{equation}

\textit{\underline{Simplified Expressions for the SIR:}} The CDF and PDF for the i.n.d. fading case can be obtained by substituting the PDF and the CDF expressions of the fully correlated case presented in Table~\ref{Tab:SIMO} in \eqref{eq:CDF_ind_SINR-SD} and \eqref{eq:PDF_SINR_SD_ind}, respectively. For i.i.d. fading, the same expressions should be substituted in \eqref{eq:CDF_iid_SINR-SD} and in \eqref{eq:PDF_SINR_SD_iid}.

\section{Performance Analysis: The Fully Correlated Case}
In this section, using the previously derived expressions for the instantaneous output SINR and focusing on the fully correlated case, important performance quality indicators are studied. The performance is evaluated using the OP and the ABEP criteria.

\subsection{Outage Probability (OP)}
The OP is defined as the probability that the SINR falls below a predetermined threshold $\gamma_{\rm th} $ and is given by $P_{\rm out}=F_{\gamma_{\rm out}}(\gamma_{\rm th})$.

\subsubsection{SNR-based Criterion}
Considering the SNR-based criterion, the OP can be obtained by using \eqref{eq:CDF_SNR_based_with_SD_IND} (for i.n.d. fading) or \eqref{eq:eq:CDF_SNR_based_with_SD_IID} (for i.i.d. fading). The corresponding CDF expression for the SIR case can be found in Table~\ref{Tab:SIMO}.
\paragraph{High SNR Approximation}The exact results presented in the previous sections do not provide a clear physical insight of the system's performance. Therefore, here, we focus on the high SNR regime in order to obtain important system-design parameters such as the diversity gain $(G_d)$ and the coding gain $(G_c)$. Additionally, these analytic expressions help to quantify the amount of performance variations, which are due to the interfering effects as well as to the receiver's architecture. At high SNR the exponential CDF can be closely approximated by $F_{\gamma_D}(x)\approx \frac{x}{\overline{\gamma}_D}$ \cite{6675883}. Based on this approximated expression, and assuming i.i.d. fading conditions (the i.n.d. case can be similarly analyzed) the CDF of $\gamma_{\rm SD}$ can be expressed as $F_{\gamma_{\rm SD}}(x)=\left[ F_{\gamma_{\rm SD}}(x)\right]^N$. Therefore following the procedure proposed in \ref{sub:PDF_fully_correlated}, and employing \cite[eq. (6.561/16)]{B:Ryzhik_book}, the OP of $\gamma_{\rm SD_{out}}$, for high SNR, can be expressed as
\begin{equation}\label{eq:OP_SNR_based_high}
F_{\gamma_{\rm SD_{out}}}(\gamma_{\rm th})\approx \underbrace{\left[\sum_{n=0}^N \binom{N}{n} \frac{\Gamma\left(L+i\right)\Gamma\left(k+i\right) }{\left(k/\overline{\gamma}_I\right)^n\Gamma\left( L\right) \Gamma\left(k\right)}\right]\gamma_{\rm th}^N}_{\mathcal{D}_1} \overline{\gamma}_D^{-N}.
\end{equation}
It is obvious that \eqref{eq:OP_SNR_based_high} is of the form $\left(G_c \overline{\gamma}_D\right)^{-G_d}$, where $G_d$ represents the diversity gain, which as expected equals $N$ and $G_c=\mathcal{D}_1^{-1/N}$ is the coding gain \cite{wang2003simple,1573870}. Therefore, the coding gain of the system is affected by the number of interferers and the average INR $(L,\overline{\gamma}_I)$, the severity of the shadowing effects on the interfering channels $k$ and the outage threshold $\gamma_{\rm th}$.

%The above result implies that, when all interferers' powers
%(i.e. PRi, PS1j and PS2k) are kept constant, interference does
%not affect the diversity order. However, when the interference
%power is growing large while the ratio of transmit powers of
%both sources versus the interferers' powers is kept constant (a
%scenario corresponding to the special case b in Section VI),
%the performance of the system cannot be improved due to the
%interference becoming dominant;

\subsubsection{SINR-based Criterion}
Considering the SINR-based criterion, the OP can be obtained by substituting \eqref{eq:CDF_SINR_fully_correlated_closed} in \eqref{eq:CDF_ind_SINR-SD} (for i.n.d. fading) or \eqref{eq:CDF_iid_SINR-SD} (for i.i.d. fading). For the SIR case, the corresponding CDF expression, which should be substituted \eqref{eq:CDF_ind_SINR-SD} (for i.n.d. fading) or \eqref{eq:CDF_iid_SINR-SD} (for i.i.d. fading), can be found in Table~\ref{Tab:SIMO}.

\paragraph{High SNR Approximation}For high values of the average SNR, by following a similar procedure as the one presented for the SNR-based case, the OP can be closely approximated as
\begin{equation}\label{eq:CDF_SINR_based_approxim}
F_{\gamma_{\rm SD_{out}}} (\gamma)\approx\underbrace{\left(1+L\overline{\gamma}_I\gamma\right)^N}_{\mathcal{D}_2} \overline{\gamma}_D^{-N}
\end{equation}
where $G_d=N, G_c=\mathcal{D}_2^{-1/N}$, while similar conclusions with the SNR-based case can be drawn for the diversity and coding gains.

\subsection{Average Bit Error Probability (ABEP)}
The ABEP will be evaluated by using the MGF and CDF based approaches as described next.

\subsubsection{SNR-Based Criterion}
Considering i.n.d. fading conditions, substituting \eqref{eq:PDF_SNR_based_with_SD_IND} in the definition of the MGF, using \cite[eq. (07.45.26.0005.01)]{E:wolfram}, and employing \cite[eq. (7.813/1)]{B:Ryzhik_book}, yields the following expression for the MGF of $\gamma_{\rm SD_{out}}$
\begin{equation}\label{eq:MGF_SNR_based_with_SD_IND_SINR}
\begin{split}
&M_{\gamma_{\rm SD_{out}}}(s)\hspace{-0.1cm}=\hspace{-0.1cm}\sum_{\substack{n,n\\ \lambda_1,\ldots,\lambda_n}}^N \hspace{-0.1cm}\frac{\left(-\mathcal{S}_{1,\overline{\gamma}_{D_{\lambda_m}}}^n\right)}{\Gamma(k)\Gamma(L)}
 \left[ \frac{k \left(\mathcal{S}_{1,\overline{\gamma}_{D_{\lambda_m}}}^n+s\right)}{\overline{\gamma}_{I} \mathcal{S}_{1,\overline{\gamma}_{D_{\lambda_m}}}^n}\right]^{ \frac{L+k-1}{2}}  \\ &  \times \hspace{-0.1cm}\left\{\hspace{-0.1cm} \left(\mathcal{S}_{1,\overline{\gamma}_{D_{\lambda_m}}}^n+s\right)^{-1} \hspace{-0.1cm} \mathcal{G} \substack{1,3
\\ 3,1} \left(  \frac{\frac{\mathcal{S}_{1,\overline{\gamma}_{D_{\lambda_m}}}^n\overline{\gamma}_I}{ k} }{ \mathcal{S}_{1,\overline{\gamma}_{D_{\lambda_m}}}^n+s}
\Bigg| \substack{\frac{L+k-1}{2},\hspace{-0.1cm}\frac{1-L+k}{2},\hspace{-0.1cm}\frac{1+L-k}{2}
\\ \null \\
\frac{L+k-1}{2}}\right) \right. \\ & \left. +\left(\mathcal{S}_{1,\overline{\gamma}_{D_{\lambda_m}}}^n\right)^{-1}\mathcal{G} \substack{1,3
\\ 3,1} \left(  \frac{\frac{\mathcal{S}_{1,\overline{\gamma}_{D_{\lambda_m}}}^n\overline{\gamma}_I}{ k} }{ \mathcal{S}_{1,\overline{\gamma}_{D_{\lambda_m}}}^n+s}
\Bigg| \substack{\frac{L+k+1}{2},\frac{1-L+k}{2},\frac{1+L-k}{2}
\\ \null \\
\frac{L+k+1}{2}}\right)\right\}.
\end{split}
\end{equation}
For the i.i.d. fading case, based on \eqref{eq:eq:PDF_SNR_based_with_SD_IID}, \eqref{eq:MGF_SNR_based_with_SD_IND_SINR} simplifies to
\begin{equation}\label{eq:MGF_SNR_based_with_SD_IID}
\begin{split}
&M_{\gamma_{\rm SD_{out}}} (s)=\sum\limits_{n=1}^N \binom{N}{n} (-1)^{n+1}   \left[ \frac{\left(n/\overline{\gamma}_D+s\right)k\overline{\gamma}_D}{n \overline{\gamma}_I}\right]^{\frac{L+k-1}{2}}   \\ & \times \frac{\left( 1+\frac{s\overline{\gamma}_D}{n}\right)}{\Gamma(k) \Gamma(L)} \left\{ \mathcal{G} \substack{1,3
\\ 3,1} \left(  \frac{n \overline{\gamma}_I/  k  }{ n+s\overline{\gamma}_D}
\bigg| \substack{\frac{L+k-1}{2},\frac{k-L+1}{2},\frac{1+L-k}{2}
\\ \null \\
\frac{L+k-1}{2}}\right) \right. \\ & \left.+ \left( 1+\frac{s\overline{\gamma}_D}{n}\right)\mathcal{G} \substack{1,3
\\ 3,1} \left(  \frac{n \overline{\gamma}_I/  k  }{ n+s\overline{\gamma}_D}
\bigg| \substack{\frac{L+k+1}{2},\frac{k-L+1}{2},\frac{1+L-k}{2}
\\ \null \\
\frac{L+k+1}{2}}\right)\right\}
\end{split}
\end{equation}
which for the SISO system becomes
\begin{equation}\label{eq:MGF_SINR_Rayleigh_desired}
\begin{split}
&M_{\gamma_{\rm out}}\left(s\right)=  \frac{\left(k \overline{\gamma}_D/\overline{\gamma}_I\right)^{\frac{L+k}2}}{\Gamma(L) \Gamma(k)\overline{\gamma}_D} \left\{ \left(\frac1{\overline{\gamma}_D}+s \right)^{\frac{L+k}2-1} \right. \\ & \left. \times  \mathcal{G} \substack{1,3
\\ 3,1} \left(  \frac{\overline{\gamma}_I/\left( k \overline{\gamma}_D\right)}{ 1/ \overline{\gamma}_D+s}
\bigg| \substack{\frac{L+k}{2},1-\frac{L-k}{2},1+\frac{L-k}{2}
\\ \null \\
\frac{L+k}{2}}\right) + \left(\frac1{\overline{\gamma}_D}+s \right)^{\frac{L+k}2} \right. \\ & \left. \times \overline{\gamma}_D \mathcal{G} \substack{1,3
\\ 3,1} \left(  \frac{\overline{\gamma}_I/\left( k \overline{\gamma}_D\right)}{ 1/ \overline{\gamma}_D+s}
\bigg| \substack{\frac{L+k}{2}+1,1-\frac{L-k}{2},1+\frac{L-k}{2}
\\ \null \\
\frac{L+k}{2}+1}\right)\right\}.
\end{split}
\end{equation}
For the SIR and for both i.n.d. and i.i.d. fading conditions, MGF expressions for $\gamma_{\rm SD_{out}}$ are included in Table~\ref{Tab:SIMO} (or Table~\ref{Tab:SISO} for the SISO system). Using the previously derived MGF expressions  and following the MGF-based approach, the ABEP can be readily evaluated for a variety of modulation schemes \cite{B:Sim_Alou_Book}. More specifically, the ABEP can be
calculated: \textit{i)} directly for non-coherent differential binary phase shift keying (DBPSK), that is
$P_{\rm{b}}^{DBPSK}=0.5\mathcal{M}_{\gamma_{\rm{out}}}(1)$; and
\textit{ii)} via numerical integration for Gray encoded $M$-PSK,
that is $P_{\rm{b}}^{M-PSK}= \frac1{\pi \log_2M}
\int\limits_0^{\pi-\pi/M} \mathcal{M}_{\gamma_{\rm out}}\left[\frac{\log_2M
\sin^2 \left( \pi/M\right)}{\sin^2\phi}\right] $ $d\phi$.

\paragraph{High SNR Approximation}
To evaluate the ABEP performance at the high SNR regime, the CDF-based approach will be employed. Specifically, the ABEP can be directly evaluated as
\begin{equation}\label{eq:BER_with_CDF_based}
P_{\rm b}=\alpha \beta \int_0^\infty \exp(-\beta \gamma) F_{\gamma_{\rm SD_{out}}} (\gamma) d\gamma
\end{equation}
where $\alpha,\beta$ are constants depending on the type of modulation, e.g., for DBPSK modulation $\alpha=1/2,\beta=1$ \cite{14_Aissa_Tellamburas_1}. Substituting \eqref{eq:OP_SNR_based_high} in \eqref{eq:BER_with_CDF_based} and using \cite[eq. (3.351/3)]{B:Ryzhik_book}, yields the following approximation for the ABEP
\begin{equation}\label{eq:ABEP_SNR_based_approximation}
P_{\rm b}\approx \underbrace{\left[\sum_{n=0}^N \binom{N}{n} \frac{\alpha\Gamma\left(L+i\right)\Gamma\left(k+i\right) \Gamma \left( N+1\right)}{\left(m/\overline{\gamma}_I\right)^n\Gamma\left( L\right) \Gamma\left(k\right) \beta^{N}}\right]}_{\mathcal{D}_3} \overline{\gamma}_D^{-N}
\end{equation}
and $G_d=N, G_c=\mathcal{D}_3^{-1/N}$.
\begin{figure}[!t]
\centering
\includegraphics[keepaspectratio,width=3.2in]{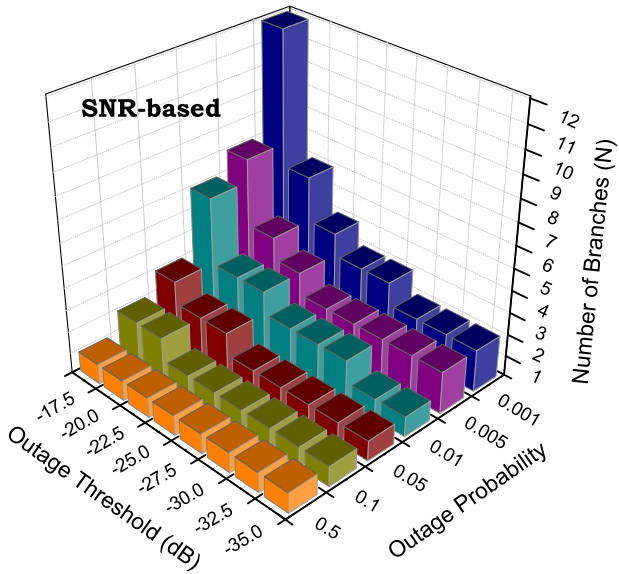}
\caption{The number of branches as a function of the OP and the normalized outage threshold for SNR-based SD reception.} \label{Fig:Fig2}
\end{figure}
\subsubsection{SINR-Based Criterion}
To evaluate the ABEP performance for the SINR-based criterion the CDF-based approach is used. For the i.n.d. fading case, \eqref{eq:CDF_ind_SINR-SD} is substituted in \eqref{eq:BER_with_CDF_based}, whilst for i.i.d. fading \eqref{eq:CDF_iid_SINR-SD} is employed. For both cases numerical integration techniques must be applied, using any of the well known mathematical software packages, since a direct derivation in terms of closed forms is not possible. Similar to the SINR analysis, the ABEP for the SIR can be evaluated using the CDF expressions for the fully correlated case provided in Table~\ref{Tab:SIMO}. Based on these expressions and substituting \eqref{eq:CDF_ind_SINR-SD} (for the i.n.d. fading) as well as \eqref{eq:CDF_iid_SINR-SD} (for the i.i.d. fading case) in \eqref{eq:BER_with_CDF_based} and employing numerical integration the ABEP can be readily evaluated.

\paragraph{High SNR Approximation}
Substituting \eqref{eq:CDF_SINR_based_approxim} in \eqref{eq:BER_with_CDF_based}, the following closed-form approximated expression can be derived for the ABEP in the high-SNR regime
\begin{equation}
P_{\rm b}\approx \underbrace{\alpha \left( \frac{1+L\overline{\gamma}_I}{\beta}\right)^N \Gamma \left( N+1\right)}_{\mathcal{D}_4}\overline{\gamma}_D^{-N}.
\end{equation}
From the last expression we get $G_d=N, G_c=\mathcal{D}_4^{-1/N}$.
\begin{figure}[!t]
\centering
\includegraphics[keepaspectratio,width=3.2in]{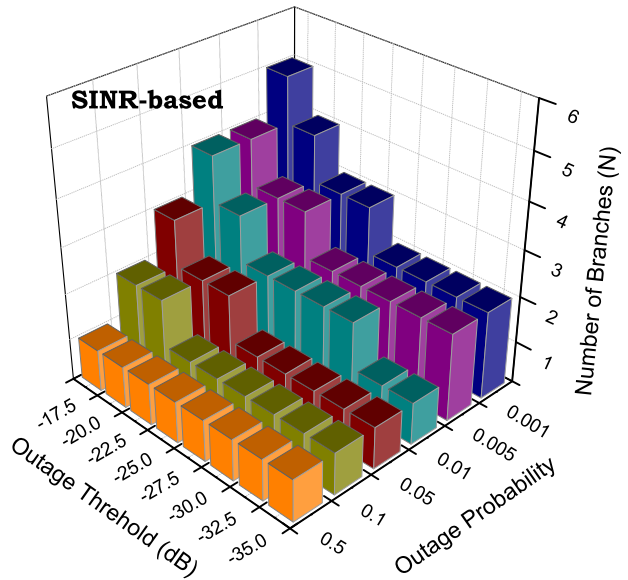}
\caption{The number of branches as a function of the OP and the normalized outage threshold for SINR-based SD reception.} \label{Fig:Fig3}
\end{figure}

\section{Numerical Results and Discussion}
In this section various numerical performance evaluation results,
which have been obtained using the previous analysis, will be presented. In particular, results related to SISO as well as SIMO systems, SNR- or SINR-based techniques and different $\mathcal{K}$-distributed i.i.d. fading and fully correlated shadowing conditions will be presented and discussed.

Considering SNR-based SD (i.e., based on \eqref{eq:eq:CDF_SNR_based_with_SD_IID}) and assuming $L=5, k=1.6, \overline{\gamma}_I=5$dB, the number of branches required for achieving a predefined target OP is plotted in Fig.~\ref{Fig:Fig2}, for different values of the normalized outage threshold, $\gamma_{\rm th}/\overline{\gamma}_{D}$. As it is shown in this figure the number of branches increases as the normalized outage threshold increases and/or the target OP decreases. Moreover, it is easily verified that for relatively low values of $\gamma_{\rm th}/\overline{\gamma}_{D}$ and/or high target OP it is not necessary to employ SD, since even with SISO the target OP is achieved. This means that the overall power consumption of the receiver side can be reduced by avoiding unnecessary circuity and channel estimations. The same observations hold also for the SINR-based SD scenario, which is depicted in Fig.~\ref{Fig:Fig3}, by utilizing \eqref{eq:CDF_iid_SINR-SD}. By comparing Figs.~\ref{Fig:Fig2}~and~\ref{Fig:Fig3} we observe that the SINR-based receiver requires considerably less reception branches than the SNR-based one for obtaining the same target OP. This gain, however, comes at the cost of much higher signal processing requirements for the SINR-based approach.
\begin{figure}
\centering
\includegraphics[keepaspectratio,width=3.2in]{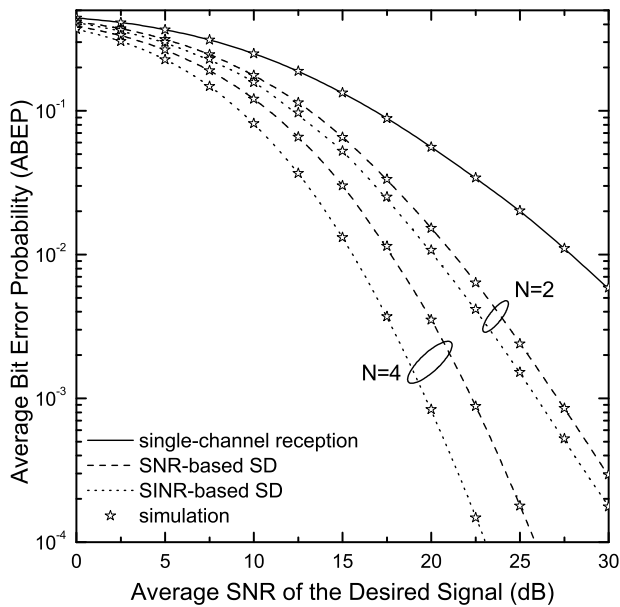}
\caption{The ABEP for SISO and SIMO SD receivers with SNR-based and SINR-based techniques, assuming DBPSK modulation.} \label{Fig:Fig4}
\end{figure}

Using \eqref{eq:MGF_SINR_Rayleigh_desired}, \eqref{eq:MGF_SNR_based_with_SD_IID} and \eqref{eq:BER_with_CDF_based}, the ABEP of DBPSK is plotted in Fig.~\ref{Fig:Fig4}, as a function of the average SNR of the desired signal, $\overline{\gamma}_D$, for SISO, SNR-based SD and SINR-based SD, respectively. In this figure the parameters are taken as $k=2, \overline{\gamma}_I=5$dB and $L=4$. As expected, the best performance is provided by the SINR-based receiver, while the performance gap between SINR and SNR-based SD increases as the number of diversity branches employed also increases.

\begin{figure}
\centering
\includegraphics[keepaspectratio,width=3.2in]{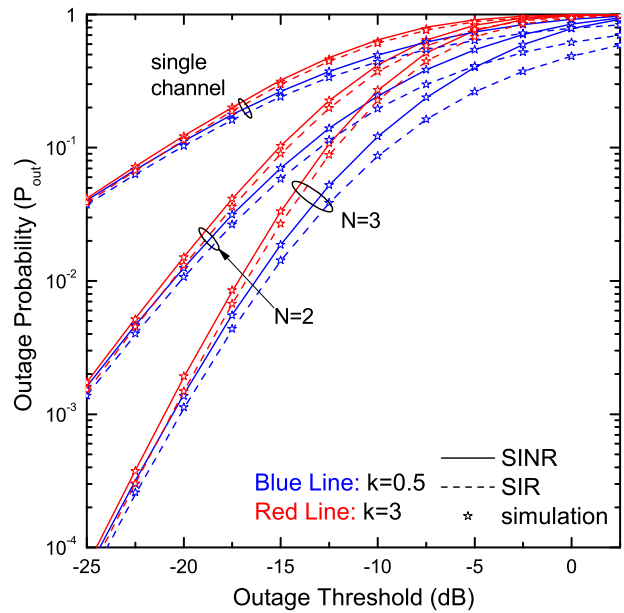}
\caption{The OP for SISO as well as SINR-based SD SIMO system.} \label{Fig:Fig5}
\end{figure}

In Fig.~\ref{Fig:Fig5}, assuming $\overline{\gamma}_I=5$dB and $L=4$, the OP is plotted as a function of the normalized outage threshold under two communication scenarios, namely SISO system and SINR-based SD, for various values of $N$. We observe that the performance improves as the number of branches increases, with a decreased rate of improvement though. A worth mentioning observation that comes out of this figure is that as the interfering signals shadowing parameter $k$ increases, the OP decreases. This is a reasonable result since severe shadowing conditions in the interfering signals result to a lower INR and thus to a higher SINR. Additionally, in order to better understand how interference affects system's performance, an interference limited communication scenario has been considered by entirely neglecting noise effects. Therefore, in the same figure, the corresponding OP performance for the SIR case has been depicted, using the CDF expression given in Table I. In all cases, when noise is not present (SIR case) the performance shows an improvement, as expected. An interesting observation though is that the performance gap between the SINR and SIR scenarios, mainly depends on the number of diversity branches employed. Specifically, the noise effects seem to play a more important role when SD reception is used with an increased number of antennas.

\begin{figure}
\centering
\includegraphics[keepaspectratio,width=3.2in]{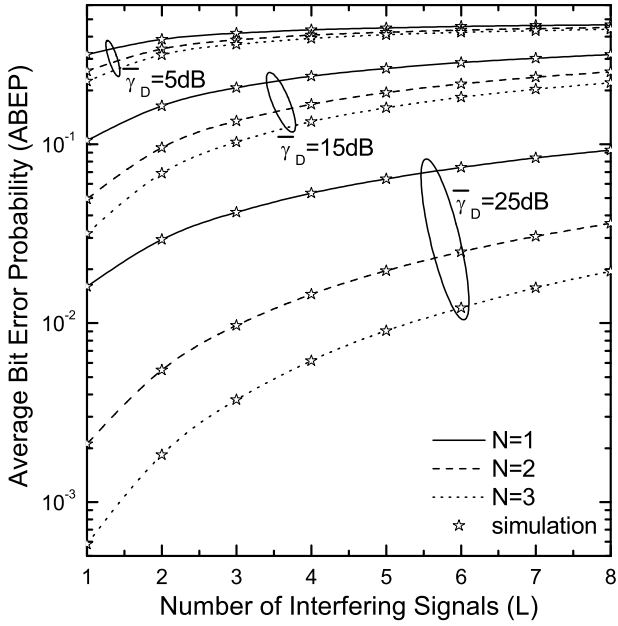}
\caption{The ABEP for SNR-based SD reception versus the number of interfering signals, assuming DBPSK modulation.} \label{Fig:Fig6}
\end{figure}

In Fig.~\ref{Fig:Fig6}, considering SNR-based SD and assuming $\overline{\gamma}_I=10$dB, $k=2$, the ABEP is plotted as a function of the number of interfering signals $L$, for various values of the number of branches $N$ and different average SNRs $\overline{\gamma}_D$ of the desired signal. We observe that as $\overline{\gamma}_D$ and/or $N$ increase the ABEP decreases, whilst in all cases the performance worsens with the increase of $L$. It is interesting to note that for higher values of $\overline{\gamma}_D$, the line gap of the performances obtained using different values of $N$ increases. For comparison purposes, computer simulation performance results are also included in Figs.~\ref{Fig:Fig4}-\ref{Fig:Fig6}, verifying in all cases the validity of the proposed theoretical approach.

\section{Conclusions}
In this paper, an analytical framework for evaluating important statistical metrics of the instantaneous output SINR of SISO as well as SIMO diversity receivers operating over composite fading channels has been presented. The proposed analysis is based on convenient expressions that have been extracted for the PDF and CDF of the sum of $\mathcal{K}$-distributed RVs, assuming identical, non-identical as well as fully-correlated distributed parameters. Focusing on the latter case, various statistical characteristics of SISO, SNR-based SD and SINR-based SD receivers are derived in closed form, which are then used to study system performance in terms of ABEP and OP. An asymptotic high SNR analysis has been also presented based on which the diversity and coding gain expressions are studied. The obtained results indicate that the combination of fading/shadowing and interference disrupts seriously the performance of the system. Furthermore, it is shown that a power efficient solution that can considerably improve this situation is the employment of SD reception with a relatively small number of diversity branches.

\renewcommand{\theequation}{A-\arabic{equation}}
\setcounter{equation}{0}
\appendices

\section{Proof of Theorem~\ref{theor:sum_independent_K_ind}}\label{App:proof_of_theoremI}
The moments generating function (MGF) of $\gamma_I\triangleq \sum_{i=1}^{L}\gamma_{I_i}$ can be expressed as \cite{957106}
\begin{equation}\label{eq:e1}
\mathcal{M}_{\gamma_{I_i}}(s)= \left( \frac{k_i}{\overline{\gamma}_{I_i}s}\right)^{k_i} \exp \left( \frac{k_i}{\overline{\gamma}_{I_i}s}\right) \Gamma \left( 1-k_i,\frac{k_i}{\overline{\gamma}_{I_i}s}\right)
\end{equation}
where $\Gamma \left( \cdot,\cdot\right)$ is the upper incomplete Gamma function \cite[eq. (8.350/2)]{B:Ryzhik_book}. The MGF of $\gamma_I$, for i.n.d. interference conditions, is given by
\begin{equation}\label{eq:f1}
\mathcal{M}_{\gamma_I}(s)=\prod_{i=1}^L\mathcal{M}_{\gamma_{I_i}}(s).
\end{equation}
Therefore, assuming non integer values for $k_i$, substituting \eqref{eq:e1} in \eqref{eq:f1} and using the infinite series representation for the incomplete gamma function, i.e., \cite[eq. (8.354/2)]{B:Ryzhik_book}, yields
\begin{equation}\label{eq:h1}
\begin{split}
\mathcal{M}_{\gamma_I}(s)&= \mathcal{G}_1 \left( \frac{1}{s}\right)^{\mathcal{S}_{k_i,1}^L}   \exp \left[ \left( \mathcal{S}_{k_i,\overline{\gamma}_{I_i}}^L\right)\frac{1}{s}\right]  \left[ \prod_{i=1}^L \left( 1-\hat{t}_{h,i}\right)\right]
\end{split}
\end{equation}
where
\begin{equation}\label{eq:t_h_i}
\begin{split}
\hat{t}_{h,i}&=\sum\limits_{h=0}^\infty \underbrace{\frac{(-1)^{h}}{h! \Gamma(1-k_i) (1-k_i+h)} \left( \frac{k_i}{\overline{\gamma}_{I_i} s}\right)^{1-k_i+h}}_{t_{h,i}}
\end{split}
\end{equation}
$\mathcal{S}_{x_q,y_q}^z=\sum\limits_{q=1}^z \frac{x_q}{y_q}$ and $\mathcal{G}_1=\left[ \prod_{i=1}^L  \left( \frac{k_i}{\overline{\gamma}_{I_i}}\right)^{k_i} \Gamma \left( 1-k_i\right)\right]$. Furthermore, since
\begin{equation}\label{Eq:product}
\begin{split}
& \prod_{i=1}^L  \left( 1-\hat{t}_{h,i} \right)=1  + \sum_{\substack{i,i\\ \lambda_1,\ldots,\lambda_i}}^L
\prod_{n=1}^i \hat{t}_{h,\lambda _n}
\end{split}
\end{equation}
where $\sum\limits_{\substack{x,y\\ \lambda_1,\ldots,\lambda_x}}^z =
\sum\limits_{x=1}^z ( -1
)^{y} \sum\limits_{\lambda_1 =1}^{z-x+1}\sum\limits_{\lambda _{2} =\lambda
_1 +1}^{z-x+2} \cdots \sum\limits_{\lambda_x =\lambda_{x-1} +1}^z$,
\eqref{eq:h1} can be rewritten as
\begin{equation}\label{eq:i1}
\begin{split}
\mathcal{M}_{\gamma_I}(s)&= \left[ \prod_{i=1}^L  \left( \frac{k_i}{\overline{\gamma}_{I_i}}\right)^{k_i} \Gamma \left( 1-k_i\right)\right] \left( \frac{1}{s}\right)^{\mathcal{S}_{k_i,1}^L}   \\ & \times \exp \left( \frac{\mathcal{S}_{k_i,\overline{\gamma}_{I_i}}^L}{s}\right)   \left( 1  + \sum_{\substack{i,i\\ \lambda_1,\ldots,\lambda_i}}^L
\prod_{n=1}^i \hat{t}_{h,\lambda _n} \right).
\end{split}
\end{equation}
Capitalizing on the power series identity $
\sum_{k=0}^\infty \alpha_kx^k \sum_{k=0}^\infty \beta_kx^k=\sum_{k=0}^\infty c_kx^k,$ where $c_h=\sum_{k=0}^h \alpha_k\beta_{h-k}, $ given in \cite[eq. (0.316)]{B:Ryzhik_book},
\eqref{eq:h1} can be simplified as
\begin{equation}\label{eq:k1}
\begin{split}
\mathcal{M}_{\gamma_I}(s)& = \left[ \prod_{i=1}^L  \left( \frac{k_i}{\overline{\gamma}_{I_i}}\right)^{k_i} \Gamma \left( 1-k_i\right)\right] \left( \frac{1}{s}\right)^{\mathcal{S}_{k_i,1}^L} \\ & \times \exp \left( \frac{\mathcal{S}_{k_i,\overline{\gamma}_{I_i}}^L}{s}\right)  \left( 1  + \sum_{\substack{i,i\\ \lambda_1,\ldots,\lambda_i}}^L
\sum\limits_{h=0}^\infty \mathcal{G}_3 \right).
\end{split}
\end{equation}
The MGF expression presented in \eqref{eq:k1} is in a convenient form for applying the inverse Laplace transform given in \cite[eq. (29.3.81)]{B:abramowitz}, leading to \eqref{eq:l1}, which completes the proof.

\renewcommand{\theequation}{B-\arabic{equation}}
\setcounter{equation}{0}

\section{Proof for the Convergence of the Infinite Series in $IS$.}\label{App:bound}
For $L=3$, term $IS$ in \eqref{eq:l1} can be expressed as
\begin{equation}\label{eq:1}
\begin{split}
IS&=\hspace{-0.08cm}-\underbrace{\sum_{\lambda_1}^3 \sum_{h=0}^\infty \hspace{-0.08cm}\left( \frac{\mathcal{S}_{k_i,\overline{\gamma}_{I_i}}^L} {\gamma}\right)^{\frac{1-\mathcal{G}_{2_1}}2} \hspace{-0.08cm} \mathcal{G}_3 I_{\mathcal{G}_{2_1}-1}\hspace{-0.05cm} \left( 2\sqrt{\mathcal{S}_{k_i,\overline{\gamma}_{I_i}}^L \gamma} \right)}_{IS_1} \\ &+ \underbrace{ \sum_{\substack{\lambda_1,\lambda_2}}^{2,3} \sum_{h=0}^\infty \left( \frac{\mathcal{S}_{k_i,\overline{\gamma}_{I_i}}^L} {\gamma}\right)^{\frac{1-\mathcal{G}_{2_2}}2}\mathcal{G}_3 I_{\mathcal{G}_{2_2}-1} \left( 2\sqrt{\mathcal{S}_{k_i,\overline{\gamma}_{I_i}}^L \gamma} \right)}_{IS_2}\\ & -\underbrace{\sum_{\substack{\lambda_1,\lambda_2,\lambda_3}}^{1,2,3} \sum_{h=0}^\infty \underbrace{\left( \frac{\mathcal{S}_{k_i,\overline{\gamma}_{I_i}}^L}{\gamma}\right)^{\frac{1-\mathcal{G}_{2_3}}2} \hspace{-0.05cm}\mathcal{G}_3 I_{\mathcal{G}_{2_3}-1} \hspace{-0.05cm}\left( 2\sqrt{\mathcal{S}_{k_i,\overline{\gamma}_{I_i}}^L \gamma} \right)}_{d_{1,h}}}_{IS_3}.
\end{split}
\end{equation}
It is well known that the sum of convergent series also converges. In the following analysis we will show that $IS_3$ converges. It is noted that the same analysis can be followed for $IS_1$ and $IS_2$, which are special cases of $IS_3$. To this end if $S_1=\sum\limits_{\substack{\lambda_1,\lambda_2,\lambda_3}}^{1,2,3} \sum\limits_{h=0}^\infty |d_{1,h}|$ converges, so is $IS_3$, since if a series is absolute convergent then it is also convergent. Substituting $d_{1,h}$ from \eqref{eq:1} and after some mathematical manipulations yields
\begin{equation}\label{eq:L3_truncation_error_2}
\begin{split}
&S_1=\sum_{\substack{\lambda_1,\lambda_2,\lambda_3}}^{1,2,3}
\mathcal{B}_1 \sum_{h=0}^\infty \sum_{q=0}^h \sum_{t=0}^q  \frac{|(-1)^h|/t!}{  |1-k_{\lambda_1}+t |} \\ & \times \frac{\left( \frac{k_{\lambda_1}}{\overline{\gamma}_{I_{\lambda_1}}}\right)^{t}\left( \frac{k_{\lambda_2}}{\overline{\gamma}_{I_{\lambda_2}}}\right)^{q-t}}{(q-t)!  |1-k_{\lambda_2}+q-t |}  \\ & \times  \frac{\left( \frac{k_{\lambda_3}}{\overline{\gamma}_{I_{\lambda_3}}}\right)^{h-q}\left(\frac{\mathcal{S}_{k_i, \overline{\gamma}_{I_i}}^L}{\gamma}\right)^{\frac{h}{2}}}{(h-q)!  |1-k_{\lambda_3}+h-q |}I_{\mathcal{G}_{2_3}-1} \left( 2\sqrt{\mathcal{S}_{k_i,\overline{\gamma}_{I_i}}^L\gamma}\right)
\end{split}
\end{equation}
where $ \mathcal{B}_1=\left[\prod\limits_{i=1}^3 \frac{\left( k_{\lambda_i}/\overline{\gamma}_{I_{\lambda_i}} \right)^{1-k_{\lambda_i}}}{|\Gamma\left(1-k_{\lambda_i}\right)|}\right] \left(\frac{\mathcal{S}_{k_i,\overline{\gamma}_{I_i}}^L}{\gamma}\right)^{\sum\limits_{r=1}^3 \frac{k_{\lambda_r}}2 - \sum\limits_{j=1}^L \frac{k_j}2-1}.$ Noticing that for any value of $h,q,t$, $\frac{|(-1)^h|}{|1-k_{\lambda_1}+t|}\leq \alpha_1$, $ \frac{1}{(q-t)! |1-k_{\lambda_2}+q-t|}\leq \alpha_2$, $\frac{1}{(h-q)! |1-k_{\lambda_3}+h-q|}\leq \alpha_3$, where $\alpha_i=\max\big( \frac{2k_{\lambda_1}}{|1-k_{\lambda_i}|},\frac{1}{r_{\lambda_i}},$ $\frac{1}{1-r_{\lambda_i}} \big)$ with $r_{\lambda_i}$ denoting the decimal part of $k_{\lambda_i}$, \eqref{eq:L3_truncation_error_2} can be written
\begin{equation}\label{eq:L3_truncation_error_3}
\begin{split}
&S_2 \hspace{-0.1cm}=\hspace{-0.1cm}\sum\limits_{\substack{\lambda_1,\lambda_2,\lambda_3}}^{1,2,3}
\hspace{-0.1cm}\underbrace{\hspace{-0.1cm}\mathcal{B}_1 \hspace{-0.1cm}\left[\prod_{i=1}^3\alpha_i\right]\hspace{-0.1cm} }_{\mathcal{B}_2}\sum_{h=0}^\infty \sum_{q=0}^h \sum_{t=0}^q\frac{\left( \frac{k_{\lambda_1}}{\overline{\gamma}_{I_{\lambda_1}}}\right)^{t}}{t!}\hspace{-0.1cm}\hspace{-0.1cm} \left( \frac{k_{\lambda_2}}{\overline{\gamma}_{I_{\lambda_2}}}\right)^{q-t}\\ & \times \hspace{-0.1cm}\left( \frac{k_{\lambda_3}}{\overline{\gamma}_{I_{\lambda_3}}}\right)^{h-q}\hspace{-0.1cm}\left( \frac{\mathcal{S}_{k_i,\overline{\gamma}_{I_i}}^L}{\gamma}\right)^{\frac{h}{2}} \hspace{-0.1cm}I_{\mathcal{G}_{2_3}-1} \left( 2\sqrt{\mathcal{S}_{k_i,\overline{\gamma}_{I_i}}^L\gamma}\right).
\end{split}
\end{equation}
Furthermore, employing the infinite series representation for the Bessel function, i.e., $I_v(z)=\sum_{j=0}^\infty \frac{\left( z/2\right)^{2k+v}}{\Gamma(j+v+1)j!} $ \cite[eq. (8.445)]{B:Ryzhik_book}, noticing that $\frac1{\Gamma\left(j+\sum_{i=1}^L k_i+3-\sum_{w=1}^3 k_{\lambda_w}+h \right)} \leq$ $ \frac{1/\Gamma\left(h+1\right)}{\Gamma\left(\sum_{i=1}^L k_i+3-\sum_{w=1}^3 k_{\lambda_w} \right) }$ and after some manipulations \eqref{eq:L3_truncation_error_3} is rewritten as
\begin{equation}\label{eq:L3_truncation_error_4}
\begin{split}
&S_3=\sum\limits_{\substack{\lambda_1,\lambda_2,\lambda_3}}^{1,2,3}
\underbrace{\mathcal{B}_2  \frac{\left(\mathcal{S}_{k_i,\overline{\gamma}_{I_i}}^L \gamma\right)^{\sum\limits_{i=1}^L \frac{k_i}2+\frac32-\sum\limits_{w=1}^3 \frac{k_{\lambda_w}}2}}{\Gamma\left(\sum\limits_{i=1}^L k_i+3-\sum\limits_{w=1}^3 k_{\lambda_w}\right)}}_{\mathcal{B}_3} \\ & \times\left[ \sum_{h=0}^\infty \frac{\left(\frac{k_{\lambda_3}\gamma}{\overline{\gamma}_{I_{\lambda_3}}}\right)^{h}}{\Gamma(1+h)}\sum_{q=0}^h \left(\frac{k_{\lambda_2}\overline{\gamma}_{I_{\lambda_3}}} {k_{\lambda_3}\overline{\gamma}_{I_{\lambda_2}}}\right)^{q} \right. \\ & \left. \times\sum_{t=0}^q \frac{\left( \frac{k_{\lambda_1}\overline{\gamma}_{I_{\lambda_2}}} {k_{\lambda_2}\overline{\gamma}_{I_{\lambda_1}}}\right)^{t}}{t!}  \sum_{j=0}^\infty  \frac{\left( \mathcal{S}_{k_i,\overline{\gamma}_{I_i}}^L \gamma\right)^j}{j!}   \right].
\end{split}
\end{equation}
Based on the fact that $\sum\limits_{t=0}^q \frac{1}{t!}\left( \frac{k_{\lambda_1}\overline{\gamma}_{I_{\lambda_2}}} {k_{\lambda_2}\overline{\gamma}_{I_{\lambda_1}}}\right)^{t}\leq \sum\limits_{t=0}^\infty \frac{1}{t!}\left( \frac{k_{\lambda_1}\overline{\gamma}_{I_{\lambda_2}}} {k_{\lambda_2}\overline{\gamma}_{I_{\lambda_1}}}\right)^{t}$ and $\sum\limits_{i=0}^\infty \frac{\gamma^i}{i!}=\exp(\gamma)$, yields
\begin{equation}\label{eq:L3_truncation_error_5}
\begin{split}
&S_4=\sum\limits_{\substack{\lambda_1,\lambda_2,\lambda_3}}^{1,2,3}
\mathcal{B}_3 \exp\left( \mathcal{S}_{k_i,\overline{\gamma}_{I_i}}^L \gamma \right) \exp\left( \frac{k_{\lambda_1}\overline{\gamma}_{I_{\lambda_2}}} {k_{\lambda_2}\overline{\gamma}_{I_{\lambda_1}}}\right) \\ & \times \left[ \sum_{h=0}^\infty \frac{\left(k_{\lambda_3}\gamma/\overline{\gamma}_{I_{\lambda_3}}\right)^{h}}{\Gamma(1+h)}\sum_{q=0}^h \left(\frac{k_{\lambda_2}\overline{\gamma}_{I_{\lambda_3}}} {k_{\lambda_3}\overline{\gamma}_{I_{\lambda_2}}}\right)^{q}
 \right].
\end{split}
\end{equation}
For $\frac{k_{\lambda_2}\overline{\gamma}_{I_{\lambda_3}}} {k_{\lambda_3}\overline{\gamma}_{I_{\lambda_2}}}\geq1$, it is easy to show that $\sum\limits_{q=0}^h \left(\frac{k_{\lambda_2}\overline{\gamma}_{I_{\lambda_3}}} {k_{\lambda_3}\overline{\gamma}_{I_{\lambda_2}}}\right)^{q} \leq(h+1) \left(\frac{k_{\lambda_2}\overline{\gamma}_{I_{\lambda_3}}} {k_{\lambda_3}\overline{\gamma}_{I_{\lambda_2}}}\right)^{h}$. Based on this inequality and employing the ratio test for the convergence of the infinite series appearing in \eqref{eq:L3_truncation_error_5}, it can be shown that $\lim\limits_{h\rightarrow\infty} \frac{t_{h+1}}{t_h}=0$, with $t_h=\frac{h+1}{\Gamma(h+1)}\left(\frac{k_{\lambda_2}\gamma} {\overline{\gamma}_{I_{\lambda_2}}}\right)^h$
and thus $S_4$ converges. For $\frac{k_{\lambda_2}\overline{\gamma}_{I_{\lambda_3}}} {k_{\lambda_3}\overline{\gamma}_{I_{\lambda_2}}}<1$, $\sum\limits_{q=0}^h \left(\frac{k_{\lambda_2}\overline{\gamma}_{I_{\lambda_3}}} {k_{\lambda_3}\overline{\gamma}_{I_{\lambda_2}}}\right)^{q}\ \leq \sum\limits_{q=0}^\infty \left(\frac{k_{\lambda_2}\overline{\gamma}_{I_{\lambda_3}}} {k_{\lambda_3}\overline{\gamma}_{I_{\lambda_2}}}\right)^{q}=\frac{k_{\lambda_3} \overline{\gamma}_{I_{\lambda_2}}}{k_{\lambda_3}\overline{\gamma}_{I_{\lambda_2}} -k_{\lambda_2}\overline{\gamma}_{I_{\lambda_3}} }$. Thus, similar to the previous analysis, it can be proved that $S_4$ converges. Therefore, since $S_1$ in \eqref{eq:L3_truncation_error_2} is bounded by the convergent series $S_4$, it also converges, meaning that $IS_3$ is absolute convergent, which completes the proof.

\renewcommand{\theequation}{C-\arabic{equation}}
\setcounter{equation}{0}

\section{Proof of Theorem~\ref{theor:sum_independent_K_iid}}\label{App:proof_of_theoremII}
Considering i.i.d. interference conditions the MGF function of $\gamma_I$ can be expressed as
\begin{equation}\label{eq:f}
\mathcal{M}_{\gamma_I}(s)=\left[\mathcal{M}_{\gamma_{I_i}} (s)\right]^L.
\end{equation}
Assuming non integer values for $k$, substituting \eqref{eq:e1} in this equation, with $\overline{\gamma}_{I_i}=\overline{\gamma}_{I}$, using also the infinite series representation for the incomplete gamma function, i.e., \cite[eq. (8.354/2)]{B:Ryzhik_book} and employing the binomial identity, \eqref{eq:f} can be written as
\begin{equation}\label{eq:h}
\begin{split}
&\mathcal{M}_{\gamma_I}(s) = \left( \frac{k}{\overline{\gamma}_Is}\right)^{Lk} \exp \left( \frac{L k}{\overline{\gamma}_Is}\right) \sum\limits_{i=0}^L \binom{L}{i}  (-1)^{i} \\ &  \times \Gamma \left(1-k \right)^{L-i} \left[ \left( \frac{k}{\overline{\gamma}_Is}\right)^{1-k}\sum\limits_{h=0}^\infty \frac{(-1)^h}{h! (1-k+h)} \left( \frac{k}{\overline{\gamma}_Is}\right)^h \right]^i.
\end{split}
\end{equation}
Using the useful power series identity provided in \cite[eq. (0.314)]{B:Ryzhik_book}, i.e., $\left( \sum_{q=0}^\infty \alpha_q x^q\right)^h=\sum_{q=0}^\infty c_qx^q$, with $c_0=\alpha_0^h, c_m=\frac1{m\alpha_0} \sum_{q=1}^m \left( qh-m+q\right)\alpha_qc_{m-q}$ for $m\geq1$, a simplified expression for \eqref{eq:h} can be derived as \begin{equation}\label{eq:i}
\begin{split}
\mathcal{M}_{\gamma_I}(s) &=  \exp \left( \frac{L k}{\overline{\gamma}_Is}\right) \sum\limits_{i=0}^L \binom{L}{i} \Gamma \left(1-k \right)^{L-i} (-1)^{i}   \\ & \times \left[ \sum\limits_{h=0}^\infty c_h \left( \frac{k}{\overline{\gamma}_Is}\right)^{\mathcal{G}_4} \right].
\end{split}
\end{equation}
Based on the convenient expression derived in \eqref{eq:i}, the inverse Laplace transform given in \cite[eq. (29.3.81)]{B:abramowitz} can be applied, and thus \eqref{eq:i2} is derived, which completes the proof.

\balance
\bibliographystyle{IEEEtran}
\bibliography{IEEEabrv,Bithas}

\end{document}